\documentclass[11pt,envcountsect,envcountsame]{llncs}

\usepackage{a4wide,amsmath,amssymb,enumerate,gastex}

\newcommand{\mcA}{{\mathcal{A}}}
\newcommand{\mcB}{{\mathcal{B}}}
\newcommand{\mcS}{\mathcal{S}}

\newcommand{\dC}{\mathsf{C}}
\newcommand{\dN}{\mathbb{N}}

\newcommand{\bin}{\mathrm{bin}}
\newcommand{\Decision}{\mathrm{FOMC}}

\newcommand{\dom}{\mathrm{dom}}

\newcommand{\FOTh}{\mathrm{FOTh}}
\newcommand{\modulo}{\mathrm{mod}}
\newcommand{\ol}{\overline}
\newcommand{\Real}{\mathrm{REAL}}
\newcommand{\ul}{\underline}
\newcommand{\val}{\mathrm{val}}
\newcommand{\rest}{\mathord\restriction}

\begin{document}

\title{Automatic structures of bounded degree revisited}
\author{Dietrich Kuske \and Markus Lohrey\thanks{The second author
    acknowledges support from the DFG-project GELO.}
  \institute{Universit\"at Leipzig, Institut f\"ur Informatik, Germany \\
    \email{\{kuske,lohrey\}@informatik.uni-leipzig.de}}}

\maketitle

\begin{abstract}
  The first-order theory of a string automatic structure is known to be
  decidable, but there are examples of string automatic structures with
  nonelementary first-order theories. We prove that the first-order
  theory of a string automatic structure of bounded degree is decidable in
  doubly exponential space (for injective automatic presentations,
  this holds even uniformly). This result is shown to be optimal since
  we also present a string automatic structure of bounded degree whose
  first-order theory is hard for {\sf 2EXPSPACE}. We prove similar results
  also for tree automatic structures. These findings close the gaps
  left open in~\cite{Loh03} by improving both, the lower and the
  upper bounds.
\end{abstract}

\section{Introduction}

The idea of an automatic structure goes back to B\"uchi and Elgot who
used finite automata to decide, e.g., Presburger
arithmetic~\cite{Elg61}. Automaton decidable theories~\cite{Hod82}
and automatic groups~\cite{EpsCHLPT92} are similar concepts. A
systematic study was initiated by Khoussainov and Nerode~\cite{KhoN95}
who also coined the name ``{\em automatic structure}'' (we prefer the term
``{\em string automatic structures}'' in this paper). In essence, a
structure is string automatic if the elements of the universe can be
represented as strings from a regular language (an element can be
represented by several strings) and every relation of the structure
can be recognized by a finite state automaton with several heads that
proceed synchronously.  String automatic structures received
increasing interest over the last
years~\cite{BluG00,KhoR01,IshKR02,BenLSS03,KhoRS03a,%
  Kus03,Bar06,KusL08a,KusL08b,Rub08,BarKR08}. One of the main
motivations for investigating string automatic structures is that
their first-order theories can be decided uniformly (i.e., the input
is a string automatic presentation and a first-order sentence). But
even the non-uniform first-order theory is far from efficient since
there exist string automatic structures with a nonelementary
first-order theory. This motivates the search for subclasses of string
automatic structures whose first-order theories are elementary. The
first such class was identified by the second author in~\cite{Loh03}
who showed that the first-order theory of every string automatic
structure of \emph{bounded degree} can be decided in triply
exponential alternating time with linearly many alternations. A
structure has bounded degree, if in its Gaifman graph, the number of
neighbors of a node is bounded by some fixed constant. The paper
\cite{Loh03} also presents a specific example of a string automatic
structure of bounded degree, where the first-order theory is hard for
doubly exponential alternating time with linearly many
alternations. Hence, an exponential gap between the upper and lower
bound remained. An upper bound of 4-fold exponential alternating time
with linearly many alternations was shown for \emph{tree automatic
  structures} (which are defined analogously to automatic structures
using tree automata) of bounded degree. Our paper~\cite{KusL08} proves
a triply exponential space bound for the first-order theory of an
injective $\omega$-automatic structure (that is defined via
B\"uchi-automata) of bounded degree. Here, injectivity means that
every element of the structure is represented by a {\em unique}
$\omega$-word from the underlying regular language.

In this paper, we achieve three goals: 
\begin{itemize}
\item We close the complexity gaps from \cite{Loh03} for
  string/tree automatic structures of bounded degree.
\item We investigate, for the first time, the complexity of the {\em
    uniform} first-order theory (where the automatic presentation is
  part of the input) of string/tree automatic structures of bounded
  degree.
\item We refine our complexity analysis using the growth function
of a structure. This function measures the size of a sphere in the 
Gaifman graph depending on the radius of the sphere. 
The growth function of a structure of bounded degree can be at most
exponential.
\end{itemize}
Our main results are the following:
\begin{itemize}
\item 
The uniform first-order theory
for injective string automatic presentations is {\sf 2EXPSPACE}-complete.
The lower bound already holds in the non-uniform setting, i.e. there
exists a string automatic structure of bounded degree with a 
{\sf 2EXPSPACE}-complete first-order theory. 
\item 
For every string automatic structure of bounded degree, where the 
growth function is polynomially bounded, the first-order
theory is in {\sf EXPSPACE}, and there exists an example with an
{\sf EXPSPACE}-complete first-order theory.
\item 
The uniform first-order theory for injective tree automatic presentations
belongs to {\sf 4EXPTIME}; the non-uniform one to {\sf 3EXPTIME} for arbitrary
tree automatic structures, and to {\sf 2EXPTIME} if the growth function is
polynomial. Our bounds for the non-uniform problem are sharp,
i.e., there are tree automatic structures of bounded degree (and
polynomial growth) with a {\sf 3EXPTIME}-complete
({\sf 2EXPTIME}-complete, resp.) first-order theory.
\end{itemize}
We conclude this paper with some results on the complexity of
first-order fragments with fixed quantifier alternation depth one or
two on string/tree automatic structures of bounded degree. 

\section{Preliminaries}

Let $\Gamma$ be a finite alphabet and $w \in \Gamma^*$ be 
a finite word over $\Gamma$. The length of $w$ is denoted by $|w|$.
We also write $\Gamma^n = \{ w \in \Gamma^* \mid n=|w|\}$.

Let us define $\exp(0, x) = x$ and $\exp(n+1,x) = 2^{\exp(n,x)}$ for
$x \in \mathbb{N}$. We assume that the reader has some basic knowledge
in complexity theory, see e.g. \cite{Pap94}.  By Savitch's theorem,
$\mathsf{NSPACE}(s(n)) \subseteq \mathsf{DSPACE}(s(n)^2)$ if $s(n) \geq
\log(n)$. Hence, we can just write $\mathsf{SPACE}(s(n)^{O(1)})$ for
either $\mathsf{NSPACE}(s(n)^{O(1)})$ or $\mathsf{DSPACE}(s(n)^{O(1)})$.
For $k \geq 1$, we denote with $k${\sf EXPSPACE} (resp. $k${\sf EXPTIME}) the
class of all problems that can be accepted in space (resp. time)
$\exp(k,n^{O(1)})$ on a deterministic Turing machine. For {\sf 1EXPSPACE} we
write just {\sf EXPSPACE}. A computational problem is called
\emph{elementary} if it belongs to $k${\sf EXPTIME} for some $k\in\dN$.

\subsection{Tree and string automata}  \label{S automata}

For our purpose it suffices to consider only tree automata on binary trees.
Let $\Gamma$ be a finite alphabet.  A \emph{finite
  binary tree} over $\Gamma$ is a mapping $t : \dom(t) \to \Gamma$,
where $\dom(t) \subseteq \{0,1\}^*$ is finite, nonempty, and satisfies
the following closure condition for all $w \in \{0,1\}^*$: if $\{w0,
w1\} \cap \dom(t) \neq \emptyset$, then also $w,w0 \in \dom(t)$.  With
$T_\Gamma$ we denote the set of all finite binary trees over~$\Gamma$.
A (top-down) \emph{tree automaton over $\Gamma$} is a tuple $A = (Q,
\Delta, q_0)$, where $Q$ is the finite set of states, $q_0 \in Q$ is
the initial state, and
\begin{equation}\label{transition relation}
\Delta \subseteq (Q \times \Gamma \times Q
\times Q) \cup (Q \times \Gamma \times Q) \cup (Q \times \Gamma)
\end{equation}
is the non-empty transition relation.  A \emph{successful run} of $A$ on a tree $t$
is a mapping $\rho : \dom(t) \to Q$ such that (i) $\rho(\varepsilon) =
q_0$ and (ii) for every $w \in \dom(t)$ with children $w0, \ldots, wi$
(thus $-1 \leq i \leq 1$) we have
$(\rho(w), t(w), \rho(w0), \ldots, \rho(wi)) \in \Delta$.  With
$L(A)$ we denote the set of all finite binary trees $t$ such that
there exists a successful run of $A$ on~$t$.  A set $L \subseteq
T_\Gamma$ is called \emph{regular} if there exists a finite tree
automaton $A$ with $L = L(A)$.

A tree $t$ with $\dom(t)\subseteq 0^*$ can be considered as a nonempty
string $t(\varepsilon)t(0)t(00)\dots t(0^{n-1})$ with
$n=|\dom(t)|$. In the same spirit, a finite \emph{string} automaton
can be defined as a tree automaton, where the transition relation
$\Delta$ in (\ref{transition relation}) satisfies $\Delta \subseteq (Q
\times \Gamma \times Q) \cup (Q \times \Gamma)$.

We will need the following well known facts on string/tree automata:
Emptiness (resp. inclusion) of the languages of string automata can be
decided in nondeterministic logarithmic space (resp. polynomial space), whereas
emptiness (resp. inclusion) of the languages of tree automata can be
decided in polynomial time (resp. exponential time), see
e.g.~\cite{tata2007}. In all four cases completeness holds.

\subsection{Structures and first-order logic}

A \emph{signature} is a finite set $\mcS$ of relational symbols, where
every symbol $r\in \mcS$ has some fixed arity~$m_r$. The notion of an
$\mcS$-structure (or model) is defined as usual in logic. Note that we
only consider relational structures. Sometimes, we will also use
constants, but in our context, a constant $c$ can be always replaced
by the unary relation $\{c\}$. Let us fix an $\mcS$-structure $\mcA =
(A, (r^\mcA)_{r\in\mcS})$, where $r^\mcA \subseteq A^{m_r}$.  To
simplify notation, we will write $a \in \mcA$ for $a \in A$.  For $B
\subseteq A$ we define the restriction ${\mcA}\rest B = (B, (r^\mcA
\cap B^{m_r})_{r\in\mcS})$.  Given further constants $a_1, \ldots, a_n
\in \mcA$, we write $(\mcA, a_1, \ldots, a_k)$ for the structure $(A,
(r^\mcA)_{r\in\mcS}, a_1, \ldots, a_k)$.  In the rest of the paper, we
will always identify a symbol $r\in\mcS$ with its interpretation
$r^\mcA$.

A {\em congruence} on the structure $\mcA = (A, (r)_{r\in\mcS})$
is an equivalence relation $\equiv$ on $A$ such that
for every $r \in \mcS$ and all $a_1,b_1, \ldots, a_{m_r}, b_{m_r} \in A$
we have: If $(a_1,\ldots, a_{m_r}) \in r$ and $a_1 \equiv b_1,\ldots,
a_{m_r} \equiv b_{m_r}$, then also $(b_1,\ldots, b_{m_r}) \in r$.
As usual, the equivalence class of $a \in A$ w.r.t. $\equiv$ is 
denoted by $[a]_\equiv$ or just $[a]$ and $A/_\equiv$ denotes the set of all 
equivalence classes. We define the {\em quotient structure} 
$\mcA/_\equiv = (A/_\equiv, (r/_\equiv)_{r\in\mcS})$,
where $r/_\equiv = \{ ([a_1], \ldots, [a_{m_r}]) \mid
(a_1,\ldots,a_{m_r}) \in r \}$.

The \emph{Gaifman-graph} $G(\mcA)$ of the $\mcS$-structure $\mcA$ is
the following symmetric graph:
\[ 
  G(\mcA) = (A, \{ (a,b) \in A \times A \mid 
   \bigvee_{r\in\mcS} \exists (a_1, \ldots, a_{m_r}) \in r \;
    \exists j,k : a_j = a, a_k=b \})\ .
\]
Thus, the set of nodes is the universe of $\mcA$ and there is an edge
between two elements, if and only if they are contained in some tuple
belonging to one of the relations of~$\mcA$.  With $d_{\mcA}(a,b)$,
where $a,b\in \mcA$, we denote the distance between $a$ and $b$ in
$G(\mcA)$, i.e., it is the length of a shortest path connecting $a$
and $b$ in $G(\mcA)$.  For $a \in \mcA$ and $d \geq 0$ we denote with
$S_{\mcA}(d, a) = \{ b \in A \mid d_{\mcA}(a,b) \leq d\}$ the
$d$-sphere around $a$. If $\mcA$ is clear from the context, then we
will omit the subscript $\mcA$. We say that the structure $\mcA$ is
\emph{locally finite} if its Gaifman graph~$G(\mcA)$ is locally finite
(i.e., every node has finitely many neighbors). Similarly, the
structure $\mcA$ has \emph{bounded degree}, if $G(\mcA)$ has bounded
degree, i.e., there exists a constant~$\delta$ such that every $a \in
A$ is adjacent to at most $\delta$ many other nodes in $G(\mcA)$; the
minimal such $\delta$ is called the \emph{degree of $\mcA$}.  For a
structure $\mcA$ of bounded degree we can define its {\em growth
  function} as the mapping $g_{\mcA} : \dN \to \dN$ with $g_{\mcA}(n)
= \max \{ |S_{\mcA}(n,a)| \mid a \in \mcA \}$. Note that if the
function $g_{\mcA}$ is not bounded then $g_{\mcA}(n) \geq n$ for all
$n \geq 1$. For us, it is more convenient to not have a bounded
function describing the growth. Therefore, we define the
\emph{normalized growth function $g'_\mcA$} by
$g'_\mcA(n)=\max\{n,g_\mcA(n)\}$. Note that $g_\mcA$ and $g'_\mcA$ are
different only in the pathological case that all connected components
of $\mcA$ contain at most $m$ elements (for some fixed $m$). Clearly,
$g'_{\mcA}(n)$ can grow at most exponentially.  We say that $\mcA$ has
{\em exponential growth} if $g'_{\mcA}(n) \in 2^{\Omega(n)}$; if
$g'_{\mcA}(n) \in n^{O(1)}$, then $\mcA$ has {\em polynomial growth}.
 
To define logical formulas, we fix a countable infinite set~$V$ of
variables, which evaluate to elements of structures.  \emph{Formulas
  over the signature~$\mcS$} (or \emph{formulas} if the the signature
is clear from the context) are constructed from the atomic formulas
$x=y$ and $r(x_1,\ldots,x_{m_r})$, where $r\in\mcS$ and
$x,y,x_1,\ldots,x_{m_r} \in V$, using the Boolean connectives $\lor$
and $\neg$ and existential quantification over variables
from~$V$. The Boolean connective $\land$ and universal quantification
can be derived from these operators in the usual way.  The
\emph{quantifier depth} of a formula~$\varphi$ is the maximal nesting
of quantifiers in~$\varphi$. The notion of a free variable is defined
as usual. A formula without free variables is called \emph{closed}.
If $\varphi(x_1, \ldots, x_m)$ is a formula with free
variables among $x_1, \ldots, x_m$ and $a_1, \ldots, a_m \in \mcA$,
then $\mcA \models \varphi(a_1, \ldots, a_m)$ means that $\varphi$
evaluates to true in $\mcA$ when the free variable $x_i$ evaluates to
$a_i$.  The \emph{first-order theory} of $\mcA$, denoted by
$\FOTh(\mcA)$, is the set of all closed formulas $\varphi$
such that $\mcA \models \varphi$.

\subsection{Structures from automata}

This section recalls string automatic and tree automatic structures and basic
results about them. Details can be found in the survey \cite{Rub08}.

\subsubsection{Tree and string automatic structures}

String automatic structures were introduced in~\cite{Hod82}, their systematic
study was later initiated by~\cite{KhoN95}.
Tree automatic structures were introduced in \cite{Blu99}, they
generalize string automatic structures. Here, we will first introduce 
tree automatic structures. String automatic structures can be 
considered as a special case of tree automatic structures.

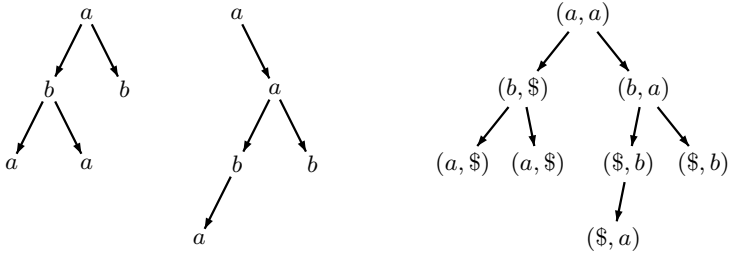
\begin{figure}[t]
  \begin{center}
    \setlength{\unitlength}{1mm}
    \begin{picture}(94,34)(0,-12)
      \gasset{linewidth=.3,Nframe=n,Nadjust=wh,Nadjustdist=0.8}
      \node(1)(10,20){$a$} \node(2)(5,10){$b$} \node(3)(15,10){$b$}
      \node(4)(0,0){$a$} \node(5)(10,0){$a$} \drawedge(1,2){}
      \drawedge(1,3){} \drawedge(2,4){} \drawedge(2,5){}
      \node(6)(30,20){$a$} \node(7)(35,10){$a$} \node(8)(30,0){$b$}
      \node(9)(40,0){$b$} \node(10)(25,-10){$a$} \drawedge(6,7){}
      \drawedge(7,8){} \drawedge(7,9){} \drawedge(8,10){}
      \node(11)(76,20){$(a,a)$} \node(12)(68,10){$(b,\$)$}
      \node(13)(84,10){$(b,a)$} \node(14)(60,0){$(a,\$)$}
      \node(15)(70,0){$(a,\$)$} \node(16)(82,0){$(\$,b)$}
      \node(17)(92,0){$(\$,b)$} \node(18)(80,-10){$(\$,a)$}
      \drawedge(11,12){} \drawedge(11,13){} \drawedge(12,14){}
      \drawedge(12,15){} \drawedge(13,16){} \drawedge(13,17){}
      \drawedge(16,18){}
    \end{picture}
  \end{center}
  \caption{The convolution of two trees}
\label{fig:convolution}
\end{figure}

Let $\Gamma$ be a finite alphabet and  
let $\$ \not \in \Gamma$ be an additional padding symbol.  Let $t_1,
\ldots, t_m \in T_\Gamma$. We define the {\em convolution} $t = t_1
\otimes \cdots \otimes t_m$, which is a finite binary tree over the
alphabet $(\Gamma \cup \{\$\})^m$, as follows: $\dom(t) =
\bigcup_{i=1}^m \dom(t_i)$ and for all $w \in \bigcup_{i=1}^m
\dom(t_i)$ we define $t(w) = (a_1, \ldots, a_m)$, where $a_i = t_i(w)$
if $w \in \dom(t_i)$ and $a_i = \$$ otherwise.  In
Fig.~\ref{fig:convolution}, the third tree is the convolution of the
first two trees.

An {\em $m$-dimensional (synchronous) tree automaton} over $\Gamma$ is just
a tree automaton $A$ over the alphabet $(\Gamma \cup \{\$\})^m$ such
that $L(A) \subseteq \{ t_1 \otimes \cdots \otimes t_n \mid t_1,
\ldots, t_m \in T_\Gamma \}$. Such an automaton defines an $m$-ary
relation
$$R(A) = \{ (t_1,\ldots, t_m) \mid  t_1 \otimes \cdots \otimes t_m \in L(A)
\}\ .$$ A {\em tree automatic presentation} is a tuple $P =
(\Gamma,A_0, A_=, (A_r)_{r \in \mcS})$, where:
\begin{itemize}
\item $\Gamma$ is a finite alphabet.
\item $\mcS$ is a signature (the signature of $P$), 
  as before $m_r$ is the arity of the symbol $r \in \mcS$.
\item $A_0$ is a tree automaton over the alphabet $\Gamma$.
\item For every $r \in \mcS$, $A_r$ is an $m_r$-dimensional tree
  automaton over the alphabet $\Gamma\cup\{\$\}$ such that $R(A_r)
  \subseteq L(A_0)^{m_r}$.
\item $A_=$ is a 2-dimensional tree automaton over the alphabet
  $\Gamma\cup\{\$\}$ such that $R(A_=) \subseteq L(A_0) \times L(A_0)$
  and $R(A_=)$ is a congruence on the structure $(L(A_0), (R(A_r))_{r
    \in {\cal S}})$.
\end{itemize}
This presentation~$P$ is called {\em injective} if $R(A_=)$ is the
identity relation on $L(A_0)$. In this case, we can omit the automaton
$A_=$ and identify $P$ with the tuple $(\Gamma,A_0, (A_r)_{r \in
  \mcS})$. The structure presented by $P$ is the quotient
$$\mcA(P) =  (L(A_0), (R(A_r))_{r \in {\cal S}})/_{R(A_=)}\ .$$ 
A structure $\mcA$ is called {\em tree automatic} if there exists a
tree automatic presentation $P$ such that $\mcA \simeq \mcA(P)$.  We
will write $[u]$ for the element $[u]_{R(A_=)}$ ($u \in L(A_0)$) of
the structure $\mcA(P)$.  We say that the presentation $P$ has bounded
degree if the structure $\mcA(P)$ has bounded degree.

A {\em string automatic presentation} is a tree automatic
presentation, where all tree automata are in fact string automata (as
explained in Section~\ref{S automata}), and a structure $\mcA$ is
called {\em string automatic} if there exists a string automatic
presentation~$P$ such that $\mcA \simeq \mcA(P)$.  Typical examples of
string automatic structures are $(\dN,+)$ (Presburger's arithmetic),
$(\mathbb{Q}, \leq)$, and all ordinals below
$\omega^\omega$~\cite{KhoN95,DelGK03}. An example of a tree automatic
structure, which is not string automatic is $(\dN, \cdot)$ (the
natural numbers with multiplication)~\cite{Blu99}, or the ordinal
$\omega^\omega$~\cite{DelGK03}. Examples of string automatic
structures of bounded degree are transition graphs of Turing machines
and Cayley-graphs of automatic groups \cite{EpsCHLPT92} (or even
right-cancellative monoids \cite{SiSt04}).

\begin{remark}
  Usually a \emph{tree automatic presentation} for an $\mcS$-structure
  $\mcA = (A, (r)_{r\in\mcS})$ is defined as a tuple $(\Gamma, L, h)$ such that
  \begin{itemize}
  \item $\Gamma$ is a finite alphabet,
  \item $L \subseteq T_\Gamma$ is a regular set of trees,
  \item $h : L \to A$ is a surjective function,
  \item the relation $\{ (u,v) \in L \times L \mid h(u)=h(v) \}$ can
    be recognized by a 2-dimensional tree automaton, and
  \item for every $r\in\mcS$, the relation $\{ (u_1, \ldots, u_{m_r})
    \in L^{m_r} \mid (h(u_1), \ldots, h(u_{m_r})) \in r \}$ can be
    recognized by an $m_r$-dimensional tree automaton.
  \end{itemize}
  Since for our considerations, tree automatic presentations are part
  of the input for algorithms, we prefer our definition, where a tree
  automatic presentation is a finite object (a tuple of finite tree
  automata), whereas in the standard definition, the presentation also
  contains the presentation map $h$.
\end{remark}

We will consider the following classes of tree automatic
presentations:
\begin{eqnarray*}
\mathsf{SA} & = & \text{the class of all string automatic presentations} \\
\mathsf{SAb} & = & \text{the class of all string automatic presentations of bounded degree} \\
\mathsf{iSAb} & = & \text{the class of all injective string automatic
  presentations of bounded degree} \\
\mathsf{TA} & = & \text{the class of all tree automatic presentations} \\
\mathsf{TAb} & = & \text{the class of all tree automatic presentations of bounded degree} \\
\mathsf{iTAb} & = & \text{the class of all injective tree automatic
  presentations of bounded degree} 
\end{eqnarray*}

\subsubsection{The model checking problem}

For the above classes of tree automatic presentations, we will be
interested in the following decision problems.

\begin{definition}
  Let $\dC$ be a class of tree automatic presentations. Then the
  \emph{first-order model checking problem $\Decision(\dC)$ for $\dC$}
  denotes the set of all pairs $(P,\varphi)$ where $P \in \dC$, and
  $\varphi$ is a closed formula over the signature of $P$ such that
  $\mcA(P) \models \varphi$.
\end{definition}

If $\dC=\{P\}$ is a singleton, then the model checking problem
$\Decision(\dC)$ for $\dC$ can be identified with the first-order
theory of the structure $\mcA(P)$.  An algorithm deciding the model
checking problem for a nontrivial class~$\dC$ decides the first-order
theories of each element of $\dC$ uniformly.

The following two results are the main motivations for investigating
tree automatic structures.

\begin{proposition}[cf.\ \cite{KhoN95,Blu99}] \label{P closure
    automatic} There exists an algorithm that computes from a tree
  automatic presentation $P = (\Gamma,A_0, A_=, (A_r)_{r \in \mcS})$
  and a formula $\varphi(x_1,\dots,x_m)$ an $m$-dimensional tree
  automaton $A$ over $\Gamma$ with $R(A) = \{ (u_1, \ldots, u_m) \in L(A_0)^m \mid
  \mcA(P) \models \varphi([u_1], \ldots, [u_m]) \}$.
\end{proposition}
The automaton is constructed by induction on the structure of the
formula~$\varphi$: disjunction corresponds to the disjoint union of
automata, existential quantification to projection, and negation to
complementation. The following result is a direct
consequence.

\begin{theorem}[cf.\ \cite{KhoN95,Blu99}]\label{T KhoNer94}  
 The model checking problem $\Decision(\mathsf{TA})$ for all tree
  automatic presentations is decidable. In particular, the first-order
  theory $\FOTh(\mcA)$ of every tree automatic structure $\mcA$ is
  decidable.
\end{theorem}
\begin{remark}
  Strictly speaking, \cite{KhoN95,Blu99} device algorithms that, given
  a tree automatic presentation and a closed formula, decide whether
  the formula holds in the presented
  structure. But a priori, it is not clear whether it is decidable,
  whether a given tuple $(\Gamma,A_0, A_=, (A_r)_{r \in \mcS})$ is a 
  tree automatic presentation. Lemma~\ref{L-TA-decidable} below shows that 
  $\mathsf{TA}$ is indeed decidable, which then completes the proof of this theorem.
\end{remark}

Theorem~\ref{T KhoNer94} holds even if we add quantifiers for ``there
are infinitely many $x$ such that~$\varphi(x)$''
\cite{Blu99,BluG00} and ``the number of elements satisfying
$\varphi(x)$ is divisible by $k$'' (for
$k\in\dN$)~\cite{KhoRS04}\footnote{\cite{KhoRS04} only provides the
  proofs for string automatic structures. These proofs are easily
  extended to tree automatic structures once the presentation is
  injective.  But every tree automatic presentation can be transformed
  into an equivalent injective one \cite[Cor.~4.2]{ColL07}.}.  This
implies in particular that it is decidable whether a tree automatic
presentation describes a locally finite structure. But the
decidability of the first-order theory is far from efficient, since
there are even string automatic structures with a nonelementary
first-order theory~\cite{BluG00}. For instance the structure
$(\{0,1\}^*, s_0, s_1, \preceq)$, where $s_i = \{ (w,wi) \mid
w \in \{0,1\}^* \}$ for $i \in \{0,1\}$ and $\preceq$ is the prefix order on
finite words, has a nonelementary first-order theory, see
e.g.~\cite[Example 8.3]{ComH90}. A locally finite example (encoding
the set of all finite labeled linearly ordered sets~\cite{Mey75}) is
as follows: the universe is the set $L=\{u\otimes v\mid
u\in\{0,1\}^+,v\in 0^*,|v|<|u|\}$. In addition, we have a partial
order $\{(u\otimes v,u\otimes v') \in L \times L \mid |v| \leq 
|v'| \}$ that encodes the union of all the linear order relations,
and a unary relation $\{u\otimes v \in L \mid\text{ position $|v|$ in
  $u$ carries }1\}$ that encodes the labeling.

\subsubsection{First complexity results: the classes $\mathsf{TA}$ etc
  and boundedness}

This paper is concerned with the uniform and non-uniform complexity of
the first-order theory of (some subclass of) tree automatic structures
of bounded degree. Thus, we will consider algorithms that take as
input tree automatic presentations (together with closed formulas).
For complexity considerations, we have to define the size $|P|$ of a
tree automatic presentation $P = (\Gamma,A_0, A_=, (A_r)_{r \in
  \mcS})$.  First, let us define the size $|A|$ of an $m$-dimensional
tree automaton $A = (Q, \Delta, q_0)$ over $\Gamma$.  A transition
tuple from $\Delta$ (see (\ref{transition relation})) can be stored
with at most $3 \log(|Q|) + m\log(|\Gamma|)$ many bits.  Hence, up to
constant factors, $\Delta$ can be stored in space $|\Delta| \cdot
(\log(|Q|) + m \log(|\Gamma|))$.  We can assume that every state is
the first component of some transition tuple, i.e., $|Q|\leq
|\Delta|$. Furthermore, the size of the basic alphabet $\Gamma$ can be
bounded by $|\Delta|$ as well, but the dimension $m$ is independent
from the size of~$\Delta$. Since our complexity measures will be up to
polynomial time reductions, it makes sense to define the size of the
tree automaton $A$ to be $|A| = |\Delta|\cdot m$. We assume $\Delta$
to be nonempty, hence $|A| \geq 1$. The size of the presentation $P =
(\Gamma,A_0, A_=, (A_r)_{r \in \mcS})$ is $|P| = |A_0|+|A_=|+\sum_{r
  \in \mcS} |A_r|$.  Note that $|\mcS| \leq |P|$ and $m \leq |P|$,
when $m$ is the maximal arity in $\mcS$.  

It will be convenient to work with injective string (resp.\ tree)
automatic presentations. The following lemma says that this is no
restriction, at least if we do not consider complexity aspects.

\begin{lemma}[\protect{\cite[Cor.~4.3]{KhoN95} and
    \cite[Cor.~4.2]{ColL07}}]\label{L-injective}
  {}From a given $P \in \mathsf{TA}$ we can compute effectively $P'
  \in \mathsf{iTA}$ with $\mcA(P) \simeq \mcA(P')$. If $P \in
  \mathsf{SA}$, then $P' \in \mathsf{iSA}$ with $\mcA(P) \simeq
  \mcA(P')$ can be computed in time $2^{O(|P|)}$.
\end{lemma}

\begin{remark}
  In \cite{ColL07}, only the existence of an equivalent injective tree
  automatic presentation is stated, but the proofs of \cite[Prop.~3.1
  and Theorem~4.1]{ColL07} are effective although the complexity is
  difficult to extract.
\end{remark}

The following lemma shows that the classes of all tree and string
automatic presentations are decidable and gives complexity
bounds. While these two results are not surprising, it is not clear
how to determine whether $\mcA(P)$ has bounded degree -- this will be
solved by Prop.~\ref{P-degree1} below.

\begin{lemma}\label{L-TA-decidable}
  The class $\mathsf{TA}$ is in~{\sf EXPTIME}, and the class
  $\mathsf{SA}$ belongs to~{\sf PSPACE}.
\end{lemma}

\begin{proof}
  We start with a proof of the first statement. Suppose we are given a
  finite alphabet~$\Gamma$, tree automata $A_0$ over $\Gamma$, and
  multi-dimensional tree automata $A_=$ and $A_r$ for $r\in\mcS$
  over~$\Gamma\cup\{\$\}$.  In a first step, we check that $L(A_=)$
  and $L(A_r)$ are languages of convolutions of elements of $L(A_0)$,
  in particular
  \begin{equation}
    \label{eq:test1}
    L(A_r)\subseteq \underbrace{L(A_0)\otimes L(A_0)\dots\otimes
    L(A_0)}_{m_r\ \text{times}}
  \end{equation}
  where $m_r$ is the arity of the automaton~$A_r$. An automaton for
  the right-hand side has size $|A_0|^{m_r}$. Thus, the inclusion
  can be decided in time exponential in $|A_r|+|A_0|^{m_r}$. Since
  $m_r$ depends on the input, this yields a doubly exponential
  algorithm. Alternatively, we proceed as follows:
  \begin{enumerate}[(a)]
  \item We check that no tree from $L(A_r)$ contains the label
    $(\$,\dots,\$)$. To this aim, replace in all transitions of $A_r$
    the letters from $(\Gamma\cup\{\$\})^{m_r}\setminus\{(\$,\dots,\$)\}$
    by $\top$ and the letter $(\$,\dots,\$)$ by $\bot$ and check
    whether the language of the resulting automaton $A_r'$ is
    contained in $T_{\{\top\}}$ (the set of all $\top$-labeled 
    binary trees). Since the set $T_{\{\top\}}$ can be
    accepted by a fixed automaton, this inclusion can be decided in
    polynomial time.
  \item Let $H\subseteq T_{\Gamma\cup\{\$\}}$ denote the set of those
    trees $t$ whose $\Gamma$-labeled nodes form an initial segment
    of~$t$ that belongs to $L(A_0)$. To accept $H$, we extend $A_0$ as
    follows (where $a \in \Gamma$):
    \begin{itemize}
    \item We add a new state $q_\$$ and transitions $(q_\$,\$)$,
      $(q_\$,\$,q_\$)$, and $(q_\$,\$,q_\$,q_\$)$.
    \item For each transition $(p,a,q)$, we add the transition
      $(p,a,q,q_\$)$.
    \item For each transition $(p,a)$, we add the transitions
      $(p,a,q_\$)$ and $(p,a,q_\$,q_\$)$.
    \end{itemize}
    Let $A_0^\$$ denote the resulting tree automaton and, for $1\le
    i\le m_r$, let $A_r^i$ denote the projection of $A_r$ to its
    $i^{th}$ component. Then we check, for all $1\le i\le m_r$ whether
    $L(A_r^i)\subseteq L(A_0^\$)$ which can be done in exponential
    time.
  \end{enumerate}
  All these tests are passed if and only if (\ref{eq:test1}) holds
  for~$A_r$. In particular, we can from now on speak of the relations
  $R(A_=)$ and $R(A_r)$ over $L(A_0)$.

  It remains to be checked that $R(A_=)$ is a congruence on the
  structure $(L(A_0),(R(A_r))_{r\in\mcS})$. For this, we proceed as
  follows
  \begin{enumerate}[(a)]
    \addtocounter{enumi}{2}
  \item First build 2-dimensional tree automata $A_\circ$, $A_{-1}$,
    and $A_{\text{id}}$ of polynomial size with
    $R(A_\circ)=R(A_=)\circ R(A_=)$, $R(A_{-1})=R(A_=)^{-1}$, and
    $R(A_{\text{id}})=\{(t,t)\mid t\in L(A_0)\}$. Then check
    $R(A_\circ)\cup R(A_{-1})\cup R(A_{\text{id}})\subseteq R(A_=)$
    which can be done in exponential time. This test is passed if and
    only if $R(A_=)$ is an equivalence relation on $L(A_0)$.
  \item For each $r\in\mcS$, first construct an $2m_r$-dimensional
    tree automaton $A_r'$ such that the tuple
    $(s_1,\dots,s_{m_r},t_1,\dots,t_{m_r})$ belongs to $R(A'_r)$ if
    and only if $(s_i,t_i)\in R(A_=)$ for all $1\le i\le m_r$ and
    $(t_1,\dots,t_{m_r})\in R(A_r)$. This can be achieved by running
    $m_r$ copies of $A_=$ as well as one copy of $A_r$ in
    parallel. Then project the automaton $A_r'$ onto the first $m_r$
    components and check whether the relation accepted by the
    resulting tree automaton is contained in~$R(A_r)$.  Although
    $A'_r$ has exponential size (since $m_r$ depends on the
    presentation~$P$), this can be done again in exponential time: we
    complement $A_r$, take the intersection with $A'_r$ and check the
    resulting automaton (of exponential size) for emptiness.
  \end{enumerate}
  This finishes the proof of the first statement. To prove the second,
  one can proceed analogously using that the inclusion problem for
  string automata belongs to {\sf PSPACE}.\qed
\end{proof}
{}From the lower bounds for inclusion of string/tree automata,
it follows easily that the upper bounds in Lemma~\ref{L-TA-decidable}
are sharp.

The following lemma says that the Gaifman graph of a string (resp.\
tree) automatic structure is effectively string (resp.\ tree)
automatic. This is an immediate consequence of Prop.~\ref{P closure
  automatic}, so the novelty lies in the estimation of the complexity.

\begin{lemma}\label{L-Gaifman-automaton}
  {}From a given tree (string) automatic presentation $P =
  (\Gamma,A_0, A_=, (A_r)_{r \in \mcS})$ one can construct a
  2-dimensional tree (string) automaton $A$ such that
  \begin{equation} \label{auto for Gaifman}
   R(A) = \{ (u,v)\in L(A_0)\times L(A_0) \mid ([u], [v])  
   \text{ is an edge of the Gaifman-graph } G(\mcA(P)) \}\ .
  \end{equation}
  If $m$ is the maximal arity in $\mcS$, then $A$ can be computed 
  in time $O(m^2\cdot |P|^2) \le |P|^{O(1)}$.
\end{lemma}

\begin{proof}
  We only give the proof for string automatic presentations, the tree
  automatic case can be shown verbatim.  Let $E$ be the edge relation
  of the Gaifman-graph $G(\mcA(P))$.  Note that for all $u,v \in
  L(A_0)$ we have $([u], [v])\in E$ iff for some $r\in\mcS$ of arity
  $m_r\le m$ and $1\le i,j\le m_r$, there exist $u_1,\dots,u_{m_r}\in
  L(A_0)$ with $(u_1,\dots,u_{m_r}) \in R(A_r)$, $u=u_i$, and
  $v=u_j$. Let $r\in\mcS$ and $1\le i,j\le m_r$.  Projecting the
  automaton $A_r$ onto the tracks $i$ and $j$, one obtains a
  2-dimensional automaton accepting all pairs $(u,v) \in\Gamma^*
  \times \Gamma^*$ such that there exists $(u_1,\dots,u_{m_r})\in
  R(A_r)$ with $u=u_i$ and $v=u_j$. Then the disjoint union of all
  these automata (for $r\in\mcS$ and $1\le i,j\le m_r$) satisfies
  (\ref{auto for Gaifman}).  Since $|\mcS|\le |P|$, the construction
  can be performed in time $O(m^2 \cdot |P|^2)$.  \qed
\end{proof}

Lemma~\ref{L-Gaifman-automaton} allows to show that also 
the bounded classes $\mathsf{TAb}$
etc.\ are decidable:

\begin{proposition}\label{P-degree1}
  The following hold:
  \begin{enumerate}[(a)] 
  \item The class $\mathsf{TAb}$ is decidable.
  \item The class $\mathsf{iTAb}$ can be decided in exponential time
    (in fact, it can be checked in polynomial time whether a given
     $P \in \mathsf{iTA}$ has bounded degree).
  \item The class $\mathsf{SAb}$ can be decided in exponential time.
  \end{enumerate}
\end{proposition}

\begin{proof}
  For statement (a), let $P \in \mathsf{TA}$ (which is decidable by
  Lemma~\ref{L-TA-decidable} in exponential time). By
  Lemma~\ref{L-injective}, we can assume $P$ to be injective. By
  Lemma~\ref{L-Gaifman-automaton} we can compute an automaton $A$ with
  (\ref{auto for Gaifman}), i.e., $A$ defines the edge relation of the
  Gaifman-graph of $\mcA(P)$.  Since $P$ was assumed to be injective
  (i.e. every equivalence class $[u]$ is the singleton $\{u\}$),
  $\mcA(P)$ is of bounded degree iff $A$ (seen as a transducer) is
  finite-valued. But this is decidable in polynomial
  time~\cite{Web90,Sei92}. This finishes the proof of~(a). 

  Next consider statement~(b): Provided the input is guaranteed to be
  an injective tree automatic presentation, the polynomial time bound
  follows from the arguments above since there is no need to apply
  Lemma~\ref{L-injective}. It remains to decide whether the input is
  indeed an injective tree automatic presentation: Using
  Lemma~\ref{L-TA-decidable}, it suffices to decide injectivity which
  can be done in exponential time by checking inclusion of $L(A_=)$ in
  the convolution of the identity on~$T_\Gamma$.

  For (c), where we start with a string automatic presentation (which
  can be decided in polynomial space and therefore exponential time by
  Lemma~\ref{L-TA-decidable}), the initial application of
  Lemma~\ref{L-injective} leads to an exponential blow-up, which gives
  in total an exponential running time for deciding the
  class~$\mathsf{SAb}$.\qed
\end{proof}

Finally, since we deal with structures of bounded degree, it will be
important to estimate the degree of such a structures given its
presentation. Such estimates are provided by the following result.

\begin{proposition}\label{P-degree}
  The following hold:
  \begin{enumerate}[(a)] 
  \item If $P \in \mathsf{iSAb}$, then the degree
    of the structure $\mcA(P)$ is bounded by $\exp(1,|P|^{O(1)})$.
  \item If $P \in \mathsf{iTAb}$, then the
     degree of the structure $\mcA(P)$ is bounded by $\exp(2,|P|^{O(1)})$.
  \item If $P \in \mathsf{SAb}$, then the degree
    of the structure $\mcA(P)$ is bounded by $\exp(2,|P|^{O(1)})$.
  \end{enumerate}
\end{proposition}

\begin{proof}
  For statement~(a) let $P \in \mathsf{iSAb}$.  {}From
  Lemma~\ref{L-Gaifman-automaton}, we can construct a string
  automaton~$A$ of size $|P|^{O(1)}$ that accepts the edge relation of
  the Gaifman graph of $\mcA(P)$. Then the degree of $\mcA(P)$ equals
  the maximal outdegree of the relation $R(A)$.  For string
  transducer, this number is exponential in the size of~$A$, i.e., it
  is in $\exp(1,|P|^{O(1)})$~\cite{Web90}.

  For (b) we can use a similar argument. But since the maximal 
  outdegree of the relation recognized by a tree transducer $A$
  is doubly exponential in the size of $A$ \cite{Sei92}, we obtain 
  the bound $\exp(2,|P|^{O(1)})$ for the degree of $\mcA(P)$.

  Finally statement (c) follows immediately from Lemma~\ref{L-injective}
  and (a).  \qed
\end{proof}

The bounds on injective string (resp. tree) automatic presentations in
Prop.~\ref{P-degree} are sharp: Let $E_n=\{(uw,vw)\mid u,v,w, \in
\{a,b\}^*, |u|=|v|=n\}$. Then the structure $(\{a,b\}^*,E_n)$ has an
injective string automatic presentation of size $O(n)$. The degree of
this structure is $2^n$. Similarly, let $E'_n$ the set of all pairs
$(t_1,t_2) \in T_{\{a,b\}} \times T_{\{a,b\}}$ of trees that differ at
most in the first $n$~levels. Then $(T_{\{a,b\}},E'_n)$ allows an
injective tree automatic presentation of size $O(n)$ and the degree of
this structure is doubly exponential in~$n$.  But it is not clear
whether the doubly exponential bound for automatic presentations in
Prop.~\ref{P-degree}(c) can be realized.  Moreover, we cannot give any
bound for general tree automatic presentations since, as already
remarked, \cite{ColL07} does not provide any estimate on the size of
an equivalent \emph{injective} tree automatic presentation.

\section{Upper bounds}

It is the aim of this section to give an algorithm that decides the
theory of a string/tree automatic structure of bounded degree. The algorithm from
Theorem~\ref{T KhoNer94} (that in particular solves this problem) is
based on Prop.~\ref{P closure automatic}, i.e., the inductive
construction of an automaton accepting all satisfying
assignments. Differently, we base our algorithm on Gaifman's
Theorem~\ref{T-Gaifman}, i.e., on the combinatorics of spheres. We
therefore start with some model theory.

\subsection{Model-theoretic background}

The following locality principle of Gaifman implies that super-exponential
distances cannot be handled in first-order logic:

\begin{theorem}[\cite{Gai82}]\label{T-Gaifman} 
  Let $\mcA$ be a structure, $(a_1,\ldots,a_k), (b_1,\ldots,b_k) \in
  \mcA^k$, $d \geq 0$, and $D_1, \ldots, D_k \geq 2^d$ such that
  \begin{equation} \label{iso}
    (\mcA \rest (\bigcup_{i=1}^k S(D_i,a_i) ), \ a_1, \ldots, a_k) \simeq
    (\mcA \rest (\bigcup_{i=1}^k S(D_i,b_i) ), \ b_1, \ldots, b_k)\ .
  \end{equation}
  Then, for every formula $\varphi(x_1,\ldots,x_k)$ of quantifier
  depth at most $d$, we have: $$\mcA\models\varphi(a_1,\ldots,a_k) \
  \Longleftrightarrow \ \mcA\models\varphi(b_1,\ldots,b_k) \ .$$
\end{theorem}
Note that (\ref{iso}) says that there is an isomorphism between the two 
induced substructures $\mcA \rest (\bigcup_{i=1}^k S(D_i,a_i) )$ and 
$\mcA \rest (\bigcup_{i=1}^k S(D_i,b_i))$ that maps $a_i$ to $b_i$ 
for all $1 \leq i \leq k$.

Let $\mcS$ be a signature and let $k,d \in \dN$ with $0 \leq k \leq d$.  A
\emph{potential $(d,k)$-sphere} is a tuple $(\mcB, b_1,\ldots, b_k)$
such that the following holds:
\begin{itemize}
\item $\mcB$ is an $\mcS$-structure with $b_1, \ldots, b_k \in
  \mcB$.
\item For all $b \in \mcB$ there exists $1 \leq i \leq k$ such that
$d_{\mcB}(b_i,b) \leq 2^{d-i}$.
\end{itemize}
There is only one $(d,0)$-sphere namely the empty sphere $\emptyset$.
For our later applications, $\mcB$ will be always a finite structure,
but in this subsection finiteness is not needed.  The potential
$(d,k)$-sphere $(\mcB, b_1,\ldots, b_k)$ is \emph{realizable in the
  structure $\mcA$} if there exist $a_1, \ldots, a_k \in \mcA$ such
that
\[
    (\mcA \rest ( \bigcup_{i=1}^k S(2^{d-i},a_i) ),a_1,\dots,a_k) 
    \simeq (\mcB, b_1,\dots,b_k)\ .
\]
Let $\sigma = (\mcB, b_1,\ldots, b_k)$ be a potential $(d,k)$-sphere
and let $\sigma' = (\mcB', b'_1,\ldots, b'_k, b'_{k+1})$ be a
potential $(d,k+1)$-sphere ($k+1 \leq d$).  
Then $\sigma'$ \emph{extends} $\sigma$
(abbreviated $\sigma \preceq \sigma'$) if 
\[
    (\mcB' \rest ( \bigcup_{i=1}^k S(2^{d-i}, b_i) ), b_1',\dots,b_k') 
    \simeq (\mcB,b_1,\dots,b_k)\ .
\]
The following definition is the basis for our decision procedure.

\begin{definition}\label{D-boolean}
  Let $\mcA$ be an $\mcS$-structure, $\psi(y_1,\ldots,y_k)$ a
  formula of quantifier depth at most~$d$, and let $\sigma =
  (\mcB, b_1,\ldots, b_k)$ be a potential $(d+k,k)$-sphere.  The
  Boolean value $\psi_\sigma \in \{0,1\}$ is defined inductively as
  follows:
  \begin{itemize}
  \item If $\psi(y_1,\ldots,y_k) $ is an atomic formula, then
    \begin{equation} \label{qf} \psi_\sigma =
      \begin{cases}
        0 & \text{if $\mcB \models \psi(b_1,\ldots,b_k)$}\\
        1 & \text{if $\mcB \not\models \psi(b_1,\ldots,b_k)$}\ .
      \end{cases}
    \end{equation}
  \item If $\psi=\neg\theta$, then $\psi_\sigma=1-\theta_\sigma$.
  \item If $\psi=\alpha\lor\beta$, then
    $\psi_\sigma=\max(\alpha_\sigma,\beta_\sigma)$.
  \item If $\psi(y_1,\ldots,y_k) = \exists y_{k+1}
    \theta(y_1,\ldots,y_k,y_{k+1})$ then
    \begin{equation} \label{existential} \psi_\sigma = \max \{
      \theta_{\sigma'} \mid \sigma' \text{ is a realizable potential
        $(d+k,k+1)$-sphere with } \sigma \preceq \sigma' \}\ .
    \end{equation}
  \end{itemize}
\end{definition}

The following result ensures for every closed formula~$\psi$ 
that $\psi_\emptyset=1$ if and only if
$\mcA\models\psi$. Hence the above
definition can possibly be used to decide validity of the
formula~$\varphi$ in the structure~$\mcA$.

\begin{proposition}\label{P upper bound}
  Let $\mcS$ be a signature, $\mcA$ an $\mcS$-structure with
  $a_1,\dots,a_k\in\mcA$, $\psi(y_1,\dots,y_k)$ a formula of
  quantifier depth at most~$d$, and $\sigma=(\mcB,b_1,\dots,b_k)$ a
  potential $(d+k,k)$-sphere with
  \begin{equation} \label{real}
    (\mcA \rest ( \bigcup_{i=1}^k S(2^{d+k-i},a_i) ), \ a_1,\ldots, a_k)
     \simeq (\mcB, b_1,\ldots,b_k)\ .
  \end{equation}
  Then $\mcA \models \psi(a_1,\ldots,a_k) \ \Longleftrightarrow \
  \psi_\sigma = 1$.
\end{proposition}

\begin{proof}
  We prove the lemma by induction on the structure of the
  formula~$\psi$.  First assume that $\psi$ is atomic, i.e. $d=0$. Then we have:
  \begin{align*}
  \psi_\sigma = 1 \ 
  & \stackrel{(\ref{qf})}{\Longleftrightarrow} \ \mcB \models \psi(b_1,\ldots,b_k) \\
  & \stackrel{(\ref{real})}{\Longleftrightarrow} \ 
      \mcA \rest ( \bigcup_{i=1}^k S(2^{k-i},a_i) ) \models
      \psi(a_1,\ldots,a_k) \\
   & \Longleftrightarrow \ \mcA \models \psi(a_1,\ldots,a_k)\ ,
  \end{align*} 
  where the last equivalence holds since $\psi$ is atomic.

  The cases $\psi=\neg\theta$ and $\psi=\alpha\lor\beta$ 
  are straightforward and therefore omitted.

  We finally consider the case $\psi(y_1,\ldots,y_k) = \exists y_{k+1}
  \theta(y_1,\ldots,y_k,y_{k+1})$.
  
  First assume that $\psi_\sigma=1$. By (\ref{existential}), there exists a realizable
  potential $(d+k,k+1)$-sphere $\sigma'$ with
  $\sigma \preceq \sigma'$ and $\theta_{\sigma'}=1$.  Since $\sigma'$ is realizable, there exist
  $a'_1, \ldots, a'_k, a'_{k+1} \in \mcA$ with
  \begin{equation} \label{real' k+1}
    (\mcA \rest ( \bigcup_{i=1}^{k+1} S(2^{d+k-i},a'_i) ), \ a'_1,\ldots, a'_k, a'_{k+1})
     \simeq (\mcB', b_1,\ldots,b_k, b_{k+1}) = \sigma'\ .
  \end{equation}
  By induction, we have $\mcA\models\theta(a_1',\ldots,a_k',a_{k+1}')$
  and therefore $\mcA\models\psi(a_1',\ldots,a_k')$. {}From
  (\ref{real}), (\ref{real' k+1}), and $\sigma\preceq\sigma'$, we also
  obtain
  $$
    (\mcA \rest ( \bigcup_{i=1}^k S(2^{d+k-i},a'_i) ), \  a'_1,\ldots, a'_k)
    \simeq
    (\mcA \rest ( \bigcup_{i=1}^k S(2^{d+k-i},a_i) ), \ a_1,\ldots, a_k)
  $$
  and therefore by Gaifman's Theorem~\ref{T-Gaifman} $\mcA \models
  \psi(a_1,\ldots, a_k)$.

  Conversely, let $a_{k+1}\in\mcA$ with
  $\mcA\models\theta(a_1,\dots,a_k,a_{k+1})$.  Let $\sigma' = (\mcB',
  b_1,\ldots,b_k,b_{k+1})$ be the unique (up to isomorphism) potential $(d+k,k+1)$-sphere
  such that
  \begin{equation} \label{real k+1}
    (\mcA \rest ( \bigcup_{i=1}^{k+1} S(2^{d+k-i},a_i) ), \ a_1,\ldots, a_k, a_{k+1})
     \simeq (\mcB', b_1,\ldots,b_k, b_{k+1})\ .
  \end{equation}
  Then (\ref{real}) implies $\sigma \preceq \sigma'$.  Moreover, by
  (\ref{real k+1}), $\sigma'$ is realizable in $\mcA$, and
  $\mcA\models\theta(a_1,\dots,a_k,a_{k+1})$ implies by induction
  $\theta_{\sigma'}=1$. Hence, by (\ref{existential}), we get $\psi_\sigma=1$ which finishes
  the proof of the lemma.\qed
\end{proof}

\subsection{The decision procedure}

Now suppose we want to decide whether the closed formula $\varphi$
holds in a tree automatic structure~$\mcA$ of {\em bounded degree}. By
Prop.~\ref{P upper bound} it suffices to compute the Boolean
value~$\varphi_\emptyset$. This computation will follow the inductive
definition of $\varphi_\sigma$ from Def.~\ref{D-boolean}. Since
every $(d,k)$-sphere that is realizable in $\mcA$ is finite, we only have 
to deal with finite spheres. The crucial part of our algorithm
is to determine whether a finite potential $(d,k)$-sphere is
realizable in~$\mcA$. In the following, for a finite potential
$(d,k)$-sphere $\sigma = (\mcB,b_1,\dots,b_k)$, 
we denote with $|\sigma|$ the number of elements of $\mcB$ and 
with $\delta(\sigma)$ we denote the degree of the finite structure
$\mcB$. We have to solve the following realizability problem:

\begin{definition}
  Let $\dC$ be a class of tree automatic presentations. Then the
  \emph{realizability problem $\Real(\dC)$ for $\dC$} denotes the set
  of all pairs $(P,\sigma)$ where
  $P \in\dC$ and $\sigma$ is a {\em finite} potential
  $(d,k)$-sphere over the signature of $P$ for some $0\le k\le d$ 
  such that $\sigma$ can be realized in $\mcA(P)$.
\end{definition}

\begin{lemma}\label{L-realisability} 
  The problems $\Real(\mathsf{iSA})$ and $\Real(\mathsf{iTA})$ are decidable.
  More precisely:
  \begin{itemize}
  \item Let $P \in \mathsf{iSA}$ and let $m$ be the maximal arity of a
    relation in $\mcA(P)$. Let $\sigma$ be a finite potential
    $(d,k)$-sphere over the signature of $P$. Then it can be checked
    in space $|\sigma|^{O(m)}\cdot |P|^2\cdot 2^{O(\delta(\sigma))}$,
    whether $\sigma$ is realizable in $\mcA(P)$.
  \item If $P \in \mathsf{iTA}$, then realizability can be checked 
    in time $\exp(1, |\sigma|^{O(m)}\cdot |P|^2\cdot 2^{O(\delta(\sigma))})$.
   \end{itemize}
\end{lemma}

\begin{proof}
  We first prove the statement on injective string automatic presentations.
  Let $P = (\Gamma,A_0, (A_r)_{r \in \mcS}) \in \mathsf{iSA}$. 
  Let $\sigma = (\mcB,b_1,\dots,b_k)$ and 
  let $c_1,\dots,c_{|\sigma|}$ be a list of all elements of $\mcB$.
  Note that every $b_i$ occurs in this list.
  Let $E_{\mcA(P)}$ be the edge
  relation of the Gaifman graph~$G(\mcA(P))$ and $E_\mcB$ that of the
  Gaifman graph~$G(\mcB)$. Then $\sigma$ is realizable
  in $\mcA(P)$ if and only if there are words $u_1,\dots,u_{|\sigma|} \in
  \Gamma^*$ such that
  \begin{enumerate}[(a)]
  \item $u_i\in L(A_0)$ for all $1\le i\le |\sigma|$,
  \item $u_i \neq u_j$ for all $1\le i<j\le |\sigma|$,
  \item $(u_{i_1},\dots, u_{i_{m_r}}) \in R(A_r)$ for all $r\in\mcS$
    and all $(c_{i_1},\dots,c_{i_{m_r}}) \in r^{\mcB}$,
  \item $(u_{i_1},\dots,u_{i_{m_r}})\notin R(A_r)$ for all $r\in\mcS$
    and all $(c_{i_1},\dots,c_{i_{m_r}})\in \mcB^{m_r}\setminus r^{\mcB}$, and
  \item there is no $v\in L(A_0)$ such that, for some $1\le j \le |\sigma|$ and
    $1\le i\le k$ with $d(c_j,b_i) < 2^{d-i}$, we have
    \begin{enumerate}[(\alph{enumi}.1)]
    \item $(u_j,v)\in E_{\mcA(P)}$ and
    \item $v \notin \{ u_p \mid (c_j,c_p)\in E_\mcB\}$.
    \end{enumerate}
  \end{enumerate}
  Then (a-d) express that the mapping $c_i\mapsto u_i$ is
  well-defined and an embedding of $\mcB$ into~$\mcA(P)$. In~(e),
  $(u_j,v)\in E_{\mcA(P)}$ implies that $v$ belongs to
  $\bigcup_{1\le i\le k}S(2^{d-i},u_i)$. Hence (e) expresses that
  all elements of $\bigcup_{1\le i\le k}S(2^{d-i},u_i)$ belong to
  the image of this embedding.

  We now construct a $|\sigma|$-dimensional automaton~$A$ over the
  alphabet $\Gamma$ that checks (a-e). At the end, we have to check
  the language of this automaton for non-emptiness. The automaton $A$
  is the direct product of automata $A_a$, $A_b$, $A_c$, $A_d$, and
  $A_e$ that check the conditions separately. Then $A_a$ is the direct
  product of $|\sigma|$ many copies of the automaton $A_0$, hence
  $A_a$ has at most $|P|^{|\sigma|}$ many states.

  Next, the automaton for (b) is the direct product of $O(|\sigma|^2)$
  many copies of an automaton of fixed size (that checks whether two
  tracks are different). Hence, this automaton has 
  $2^{|\sigma|^{O(1)}}$ many states.

  The automaton $A_c$ is again a direct product, this time of one 
  automaton for each $r\in\mcS$ (and therefore of at most~$|P|$
  many automata). Each of these automata is the direct product of
  $|r^{\mcB}|$~many copies of the automaton $A_r$. Since
  the arity of $r\in\mcS$ is bounded by $m$, we have $|r^{\mcB}| \leq
  |\sigma|^m$. Hence, the automaton $A_c$ has at most
  $(|P|^{|\sigma|^m})^{|P|} = |P|^{|P|\cdot |\sigma|^m}$ many states. 
  For~$A_d$, we can argue similarly, but this time using
  copies of the complement of the automaton~$A_r$. 
  This yields for $A_d$ the bound $(2^{|P|})^{|P|\cdot |\sigma|^m} = 
  \exp(1, |P|^2 \cdot |\sigma|^m)$ on the number of states.

  It remains to construct the automaton~$A_e$. For this, we first
  construct its complement, i.e., an automaton~$A_e'$ that checks for
  the existence of $v\in L(A_0)$ with the desired
  properties. This automaton $A_e'$ is the disjoint union of at most
  $|\sigma|$~many automata, one for each $1\le j \le |\sigma|$ such that there exists
  $1\le i\le k$ with $d(c_j,b_i)<2^{d-i}$. Any of these components
  consists of the direct product of automata $A_{e.1}$ and $A_{e.2}$
  checking (e.1) and (e.2), respectively. By
  Lemma~\ref{L-Gaifman-automaton}, $A_{e.1}$ hast at most
  $m^2\cdot |P|^2$ many states. Recall that the
  degree of $\mcB$ is $\delta(\sigma)$. Hence, the set 
  $\{ u_p \mid (c_j,c_p)\in E_\mcB\}$ 
  contains at most $\delta(\sigma)$ many elements. Thus, (e.2) can be checked by an
  automaton~$A_{e.2}$ with $2^{O(\delta(\sigma))}$ many states.  Hence, $A_e'$ is the
  disjoint union of at most $|\sigma|$ copies of an automaton of size
  $|P|^2\cdot m^2\cdot 2^{O(\delta(\sigma))}$ and therefore has
  at most $|\sigma|\cdot |P|^2\cdot m^2\cdot 2^{O(\delta(\sigma))}$ many states. 
  Now the number of states of $A_e$ can be bound
  by $\exp(1,|\sigma|\cdot |P|^2\cdot m^2\cdot 2^{O(\delta(\sigma))})$.

  In summary, the automaton $A$ has at most
  \[
    |P|^{|\sigma|}\cdot 
    2^{|\sigma|^{O(1)}}\cdot
    |P|^{|P|\cdot |\sigma|^m}\cdot
    2^{|P|^2\cdot |\sigma|^m}\cdot
    2^{|\sigma|\cdot |P|^2\cdot m^2\cdot 2^{O(\delta(\sigma))}}
    \le \exp(1,|\sigma|^{O(m)}\cdot |P|^2\cdot 2^{O(\delta(\sigma))})
  \]
  many states. Hence checking emptiness of its language (and therefore
  realizability of $\sigma$ in $\mcA(P)$) can be done in space 
  logarithmic to the number of states, i.e., in 
  space $|\sigma|^{O(m)}\cdot |P|^2\cdot 2^{O(\delta(\sigma))}$ which proves the
  statement for string automatic presentations.

  For injective tree automatic presentations, the construction and
  size estimate for~$A$ are the same as above. But emptiness of tree
  automata can only be checked in deterministic polynomial time (and
  not in logspace unless $\mathsf{NL} = \mathsf{P}$). Hence, emptiness of $A$ can be
  checked in time $\exp(1,|\sigma|^{O(m)}\cdot |P|^2\cdot
  2^{O(\delta(\sigma))})$. \qed
\end{proof}

In the following, for a tree automatic presentation $P$ of bounded
degree, we denote with $g'_P = g'_{\mcA(P)}$ the normalized growth
function of the structure $\mcA(P)$.

\begin{theorem}\label{T-upper}
  The model checking problem $\Decision(\mathsf{TAb})$ is decidable,
  i.e., on input of a tree automatic presentation~$P$ of bounded
  degree and a closed formula~$\varphi$ over the signature of~$P$, one
  can effectively determine whether $\mcA(P)\models\varphi$ holds.
  More precisely (where $m$ is the maximal arity of a relation from
  the signature of~$P$):
  \begin{enumerate}[(1)]
  \item $\Decision(\mathsf{iSAb)}$ can be decided in space
    \[
       g'_P(2^{|\varphi|})^{O(m)}\cdot \exp(2,|P|^{O(1)})
            \le \exp(2,|P|^{O(1)}+|\varphi|)\ .
    \]
  \item $\Decision(\mathsf{SAb})$ can be decided in space
    \[
       \exp(3, O(|P|)+\log(|\varphi|))\ .
    \]
  \item $\Decision(\mathsf{iTAb})$ can be decided in time
  \[
     \exp\biggl(1,\ g'_P(2^{|\varphi|})^{O(m)} \cdot \exp(3,|P|^{O(1)}) \biggr)
      \leq \exp(4, |P|^{O(1)}+\log(|\varphi|))\ .
  \]
 \end{enumerate}
\end{theorem}

\begin{proof}
  The decidability follows immediately from Theorem~\ref{T KhoNer94}
  and Prop.~\ref{P-degree1}(a).

  We first give the proof for injective string automatic presentations.
  By Prop.~\ref{P upper bound} it suffices to compute the Boolean
  value $\varphi_\emptyset$. Recall the inductive definition of
  $\varphi_\sigma$ from Def.~\ref{D-boolean} that we now translated
  into an algorithm for computing~$\varphi_\emptyset$.

  First note that such an algorithm has to handle potential
  $(d,k)$-spheres for $1 \leq k \leq d \leq|\varphi|$ ($d$ is the
  quantifier rank of $\varphi$) that are realizable in~$\mcA(P)$.  The
  number of nodes of a potential $(d,k)$-sphere realizable in
  $\mcA(P)$ is bounded by $k \cdot g'_P(2^d) \in g'_P(2^d)^{O(1)}$
  since $k\le d<2^d\le g'_P(2^d)$. The number of relations of
  $\mcA(P)$ is bounded by~$|P|$. Hence, any potential $(d,k)$-sphere
  can be described by $|P| \cdot g'_P(2^d)^{O(m)}$ many bits.

  Note that the set of $(d,k)$-spheres with $0\le k\le d$ (ordered by
  the extension relation~$\preceq$) forms a tree of depth~$d+1$. The
  algorithm visits the nodes of this tree in a depth-first manner (and
  descents when unraveling an existential quantifier). Hence we have
  to store $d+1$ many spheres. For this, the algorithm needs space
  $(d+1)\cdot |P|\cdot g'_P(2^d)^{O(m)} = |P|\cdot g'_P(2^d)^{O(m)}$.

  Moreover, during the unraveling of a quantifier, the algorithm has
  to check realizability of a potential $(d,k)$-sphere for $1\le k\le
  d\le|\varphi|$. Any such sphere has at most $g'_P(2^d)^{O(1)}$ many
  elements and the degree $\delta$ of $\mcA$ is bounded by $\exp(1,
  |P|^{O(1)})$ by Prop.~\ref{P-degree}. Hence, by
  Lemma~\ref{L-realisability}, realizability can be checked in
  space $g'_P(2^d)^{O(m)}\cdot |P|^2\cdot \exp(2,|P|^{O(1)}) \le
  g'_P(2^{|\varphi|})^{O(m)}\cdot \exp(2,|P|^{O(1)})$.

  At the end, we have to check whether a tuple $\overline{b}$
  satisfies an atomic formula $\psi(\overline{y})$, which is trivial.
  In total, the algorithm runs in space
  \begin{equation*}
     |P|\cdot g'_P(2^{|\varphi|})^{O(m)} +
     g'_P(2^{|\varphi|})^{O(m)}\cdot\exp(2,|P|^{O(1)}) 
     \le g'_P(2^{|\varphi|})^{O(m)}\cdot\exp(2,|P|^{O(1)})\ .
  \end{equation*}

  Recall that $g'_\mcA(2^{|\varphi|})\le\delta^{2^{|\varphi|}}$ and
  $\delta\le 2^{|P|^{O(1)}}$ by Prop.~\ref{P-degree}. Since also $m \le
  |P|$, we obtain
  \begin{align*}
    g'_P(2^{|\varphi|})^{O(m)}\cdot\exp(2,|P|^{O(1)})
     & \le \exp(1, |P|^{O(1)} \cdot 2^{|\varphi|} \cdot O(m)) \cdot\exp(2,|P|^{O(1)})\\
     & \le \exp(2,|P|^{O(1)}+|\varphi|)\ .
  \end{align*}
  This completes the consideration for injective string automatic
  presentations. 

  If $P$ is just automatic, we can transform it into an
  equivalent injective automatic presentation which increases the size
  exponentially by Lemma~\ref{L-injective}. Hence, replacing 
  $|P|$ by $2^{O(|P|)}$ yields the space
  bound.

  Next, we consider injective tree automatic presentations. The
  algorithm is the same, i.e., it parses the tree of all potential
  $(d,k)$-spheres and checks them for realizability. Note that the
  number of potential $(d,k)$-spheres is in $\exp(1,|P|\cdot
  g'_P(2^d)^{O(m)})$.  By Prop.~\ref{P-degree}, the degree $\delta$ is
  bounded by $\exp(2,|P|^{O(1)})$. Hence, by
  Lemma~\ref{L-realisability}, the realizability of any potential
  $(d,k)$-sphere can be checked in time
  $$\exp\biggl(1,\ g'_P(2^d)^{O(m)}\cdot |P|^2\cdot \exp(3,|P|^{O(1)}) \biggr)\le
  \exp\biggl(1,\ g'_P(2^{|\varphi|})^{O(m)} \cdot
  \exp(3,|P|^{O(1)})\biggr)\ .$$
  Recall that $g'_P(2^{|\varphi|})\le\delta^{2^{|\varphi|}}$ and
  $\delta\le \exp(2,|P|^{O(1)})$ by Prop.~\ref{P-degree}. Since also
  $m\le |P|$, we obtain
  \begin{align*}
    g'_P(2^{|\varphi|})^{O(m)} \cdot \exp(3,|P|^{O(1)})
      &\le \exp(2,|P|^{O(1)})^{2^{|\varphi|}\cdot O(|P|)}\cdot
            \exp(3,|P|^{O(1)}) \\
      &=\exp(2,{|P|^{O(1)}}+|\varphi|) \cdot \exp(3,|P|^{O(1)})\\
      &=\exp(3,|P|^{O(1)}+\log(|\varphi|))\ .
  \end{align*}\qed
\end{proof}

\begin{remark}
  Note that the above theorem does not give the complexity
  for~$\Decision(\mathsf{TAb})$, i.e., for arbitrary tree automatic
  presentations of bounded degree: Here, one can proceed as for string
  automatic presentations, i.e., make the presentation injective and
  refer to the above result on $\Decision(\mathsf{iTAb})$ -- this
  gives the decidability that we already know from Theorem~\ref{T
    KhoNer94} and Prop.~\ref{P-degree1}.  At present, we cannot
  compare the complexity of this new algorithm with the nonelementary
  one from Theorem~\ref{T KhoNer94} since the size of the injective
  presentation is not known.
\end{remark}

We derive a number of consequences on the uniform and non-uniform
complexity of the first-order theories of string/tree
automatic structures of bounded degree. The first one concerns the
uniform model checking problems and is a direct consequence of the
above theorem.

\begin{corollary}
 The following holds:
\begin{itemize}
\item The model checking problem $\Decision(\mathsf{iSAb})$ belongs to
  {\sf 2EXPSPACE}.
\item The model checking problem $\Decision(\mathsf{SAb})$ belongs to
  {\sf 3EXPSPACE}.
\item The model checking problem $\Decision(\mathsf{iTAb})$ belongs to
  {\sf 4EXPTIME}.
\end{itemize}
\end{corollary}

Next we concentrate on the non-uniform complexity, where the 
structure is fixed. For string automatic
structures, we do not get a better upper bound in this case
(statement~(i) below) except in case of polynomial growth
(statement~(ii) below).

\begin{corollary}\label{C automatic}
  Let $\mcA$ be a string automatic structure of bounded degree.
  \begin{enumerate}[(i)]
  \item Then $\FOTh(\mcA)$ belongs to {\sf 2EXPSPACE}.
  \item If $\mcA$ has polynomial growth then $\FOTh(\mcA)$ belongs to
    {\sf EXPSPACE}.
  \end{enumerate}
\end{corollary}

\begin{proof}
  Since $\mcA$ is string automatic, it has a fixed injective string
  automatic presentation~$P$, i.e., $|P|$ and $m$ are fixed constants.
  Hence the result follows immediately from (1) in
  Theorem~\ref{T-upper}.

  Now suppose that $\mcA$ has polynomial growth, i.e.,
  $g'_\mcA(x)\in x^{O(1)}$. Then, again, the claim follows
  immediately from (1) in Theorem~\ref{T-upper}, since
  $g'_\mcA(2^{|\varphi|})^{O(m)}\le 2^{O(|\varphi|)}$.\qed
\end{proof}

The last consequence of Theorem~\ref{T-upper} concerns tree automatic
structures. Here, we can improve the upper bound from
Theorem~\ref{T-upper} for the non-uniform case by one exponent. In
case of polynomial growth, we can save yet another exponent:

\begin{corollary}\label{C tree automatic}
  Let $\mcA$ be a tree automatic structure of bounded degree.
  \begin{enumerate}[(i)]
  \item Then $\FOTh(\mcA)$ belongs to {\sf 3EXPTIME}.
  \item If $\mcA$ has polynomial growth then $\FOTh(\mcA)$ belongs to
    {\sf 2EXPTIME}.
  \end{enumerate}
\end{corollary}

\begin{proof}
  Since $\mcA$ is tree automatic, it has a fixed injective
  tree automatic presentation~$P$. Hence, again, the first
  claim follows immediately from (3) in Theorem~\ref{T-upper}.

  Now suppose that $\mcA$ has polynomial growth, i.e.,
  $g'_\mcA(x)\in x^{O(1)}$. Then the claim follows since
  \[
     \exp(1,g'_\mcA(2^{|\varphi|})^{O(m)})
       \le \exp(1,2^{O(|\varphi|)}) = \exp(2,O(|\varphi|))\ ,
  \]
  implying that the problem belongs to {\sf 2EXPTIME}.\qed
\end{proof}

\subsubsection{Two observations on the growth function}

We complement this section with a short excursion into the field of
growth functions of automatic structures. The two results to be
reported indicate that these growth functions do not behave as nicely
as one would wish. Fortunately, these negative findings are of no
importance to our main concerns.

Recall that the growth rate of a regular language is either bounded by
a polynomial from above or by an exponential function from below and
that it is decidable which of these cases applies. The next lemmas
show that the analogous statements for growth functions of string
automatic structures are false.

\begin{lemma}
  There is a string automatic graph of intermediate growth (i.e., the growth
  is neither exponential nor polynomial).
\end{lemma}

\begin{proof}
  Let $L=\{0,1\}^*\$\{0,1\}^*$ and let $E$ be 
  \[
    \{(u\$bv,ub\$v)\mid u,v\in \{0,1\}^*,b\in\{0,1\}\}
    \cup \{(u\$,\$ub)\mid u\in\{0,1\}^*,b\in\{0,1\}\}\ .
  \]
  Then $T = (L,E)$ is a string automatic tree obtained from the complete
  binary tree $\$\{0,1\}^*$ by adding a path of length $n$ between $u$
  and $ub$ for $u\in\{0,1\}^n$ and $b\in\{0,1\}$.  Hence, a path of
  length $n$ starting in the root $\$$ of $T$ branches at distance
  $0,2,5,10,\ldots, i^2+1, \ldots, \lfloor\sqrt{n-1}\rfloor^2+1$ from
  the root.  Hence, for the growth function $g_T$ we
  obtain the following estimate:
  $$
  g_T(n) \in \sum_{i=0}^{\Theta(\sqrt{n})} (i+1) \cdot 2^i = \Theta(\sqrt{n})
  \cdot 2^{\Theta(\sqrt{n})} = 2^{\Theta(\sqrt{n})}
  $$
\qed
\end{proof}

\begin{lemma}
  It is undecidable whether a string automatic graph of bounded degree has
  polynomial growth.
\end{lemma}

\begin{proof}
  We show the undecidability by a reduction of the halting problem
  (with empty input) for  Turing machines. So let $N$ be a Turing
  machine. We can transform $N$ into a deterministic reversible Turing
  machine $M$ such that:
  \begin{enumerate}[(i)]
  \item $N$ halts on empty input if and only if $M$ does so.
  \item $M$ does not allow infinite sequences of backwards steps
    (i.e., there are no configurations $c_i$ with $c_{i+1}\vdash_M
    c_i$ for all $i\in\dN$), see also \cite{KusL08b} for a similar construction.
  \end{enumerate}
  Let $C$ be the set of configurations of $M$ (a regular set) and
  $c_0$ the initial configuration with empty input. Now define
  $L=(\{0,1\}C)^+$ (we assume that $0$ and $1$ do not belong to 
  the alphabet of $C$) and
  \begin{align*}
    E \ = \ & \{(uac,uac')\mid u\in L\cup\{\varepsilon\}, a\in\{0,1\}, 
                             c,c'\in C, c\vdash_M c'\} \ \cup \\
            & \{(uac,uacbc_0)\mid u\in L\cup\{\varepsilon\}, a,b\in\{0,1\},
                             c\in C\text{ is halting}\}\ .
  \end{align*}
  Then $(L,E)$ is an automatic directed graph. Since $M$ is
  reversible, it is a forest of rooted trees (by (ii)). 

  First suppose there are configurations $c_1,c_2,\dots,c_n$ with
  $c_{i-1}\vdash_M c_i$ for $1\leq  i \leq n$ such that $c_n$ is
  halting. Then the set $0(c_n\{0,1\})^*\{c_0,c_1,\dots,c_n\}$
  forms an infinite tree in $(L,E)$. Any branch in this tree branches
  every $n$ steps. Hence $(L,E)$ has exponential growth.

  Now assume that $c_0$ is the starting point of an infinite
  computation.  Let $T$ be any tree in the forest $(L,E)$. Then its
  root is of the form $uac\in L$ with $u \in L \cup \{\varepsilon\}$, 
  $a\in\{0,1\}$, and $c\in C$ such
  that $c$ is no successor configuration of any other
  configuration. There are two possibilities:
  \begin{enumerate}
  \item The configuration $c$ is the starting configuration of an
    infinite computation of $M$. Then $T$ is an infinite path.
  \item There is a halting configuration $c'$ and $n\in\dN$ with
    $c\vdash^n_M c'$. Then $T$ starts with a path of length $n$. The
    final node of this path has two children, namely $u a c' 0 c_0$
    and $u a c' 1 c_0$. But,
    since $M$ does not halt on the empty input,
    each of these nodes is the root of an infinite path.
  \end{enumerate}
  Thus, in this case $(L,E)$ has polynomial (even linear) growth.\qed
\end{proof}

\section{Lower bounds}

In this section, we will prove that the upper complexity bounds for
the non-uniform problems (Cor.~\ref{C automatic} and
Cor.~\ref{C tree automatic}) are sharp. This will imply that the complexity of
the uniform problem for injective string automatic presentations from
Theorem~\ref{T-upper} is sharp as well.

For a binary relation $r$ and $m \in \dN$ we denote with $r^m$ the
$m$-fold composition of $r$. Then the following lemma is folklore.

\begin{lemma} \label{L long paths}
  Let the signature $\mcS$ contain a binary symbol $r$.  {}From a given
  number $m$ (encoded unary), we can construct in linear time a
  formula $\varphi_m(x,y)$ such that for every
  $\mcS$-structure $\mcA$ and all elements $a,b \in \mcA$ we have:
  $(a,b) \in r^{2^m}$ if and only if $\mcA \models \varphi_m(a,b)$.
\end{lemma}
\begin{proof}
  Let $\varphi_0(x,y)=r(x,y)$ and, for $m>0$ define
  \[
    \varphi_m(x,y)=\exists
    z\forall x',y'(((x'=x\land y'=z)\lor(x'=z\land y'=y))
     \rightarrow\varphi_{m-1}(x',y'))\ .
  \]
  \qed
\end{proof}

For a bit string $u = a_1 \cdots a_m$ ($a_i \in \{0,1\}$) let $\val(u)
= \sum_{i=0}^{m-1} a_{i+1} 2^i$ be the integer value represented by
$u$. Vice versa, for $0 \leq i \leq 2^m-1$ let $\bin_m(i) \in
\{0,1\}^m$ be the unique string with $\val(\bin_m(i)) = i$.

\begin{theorem} \label{T 2EXPSPACE}
There exists a fixed string automatic structure $\mcA$ of bounded degree
such that $\FOTh(\mcA)$ is {\sf 2EXPSPACE}-hard.
\end{theorem}

\begin{proof}
  Let $M$ be a fixed Turing machine with a space bound of $\exp(2,n)$
  such that $M$ accepts a {\sf 2EXPSPACE}-complete language; such a machine
  exists by standard arguments.  Let $\Gamma$ be the tape alphabet,
  $\Sigma\subseteq \Gamma$ be the input alphabet, and $Q$ be the set
  of states. The initial (resp. accepting) state is $q_0 \in Q$
  (resp. $q_f \in Q$), the blank symbol is $\Box \in \Gamma \setminus
  \Sigma$. Let $\Omega = Q \cup \Gamma$.  A configuration of $M$ is
  described by a string from $\Gamma^* Q \Gamma^+ \subseteq \Omega^+$
  (later, symbols of configurations will be preceded with additional
  counters). For two configurations $u$ and $v$ with $|u|=|v|$ we
  write $u \vdash_M v$ if $u$ can evolve with a single $M$-transition
  into $v$.  Note that there exists a relation $\alpha_M \subseteq
  \Omega^3 \times \Omega^3$ such that for all configurations $u = a_1
  \cdots a_m$ and $v = b_1 \cdots b_m$ ($a_i, b_i \in \Omega$) we have
  \begin{equation}\label{equiv alpha}
  u \vdash_M v \quad \Longleftrightarrow \quad
  \forall i \in \{1, \ldots,m-2\} : (a_i a_{i+1} a_{i+2}, b_i b_{i+1}
  b_{i+2}) \in \alpha_M.
  \end{equation}
  Let $\Delta = \{0,1,\#\} \cup \Omega$, and let $\pi : \Delta \to
  \Omega \cup \{\#\}$ be the projection morphism with $\pi(a)=a$ for
  $a \in \Omega \cup \{\#\}$ and $\pi(0)=\pi(1)=\varepsilon$.  For $m
  \in \dN$, a string $x \in \Delta^*$ is an \emph{accepting
    $2^m$-computation} if $x$ can be factorized as $x = x_1 \# x_2 \#
  \cdots x_n\#$ for some $n \geq 1$ such that the following holds:
  \begin{itemize}
  \item For every $1 \leq i \leq n$ there exist 
  $a_{i,0}, \ldots, a_{i,2^m-1} \in \Omega$ 
  such that $x_i = \prod_{j=0}^{2^m-1} \bin_m(j) a_{i,j}$.
  \item For every $1 \leq i \leq n$,  $\pi(x_i) \in \Gamma^* Q \Gamma^+$.
  \item $\pi(x_1) \in q_0 \Sigma^* \Box^*$ and 
  $\pi(x_n) \in \Gamma^* q_f \Gamma^+$
  \item For every $1 \leq i < n$,  $\pi(x_i) \vdash_M \pi(x_{i+1})$.
  \end{itemize} 
  {}From $M$ we now construct a fixed string automatic structure $\mcA$ of
  bounded degree. We start with the following regular language $U_0$:
  \begin{align}
  U_0 \ =\ & \pi^{-1}( (\Gamma^* Q \Gamma^+\#)^*) \ \cap \label{U0-1}\\
        & (0^+ \Omega ( \{0,1\}^+ \Omega )^* 1^+ \Omega \#)^+  \ \cap \label{U0-2}\\
        & 0^+ q_0 (\{0,1\}^+ \Sigma)^* (\{0,1\}^+ \Box)^* \# \Delta^* \ \cap
        \label{U0-3}\\
        & \Delta^* q_f (\Delta \setminus \{\#\})^* \# \label{U0-4}
  \end{align}
  A string $x \in U_0$ is a candidate for an accepting
  $2^m$-computation of $M$.  With (\ref{U0-1}) we describe the basic
  structure of such a computation, it consists of a list of
  configurations separated by~$\#$.  Moreover, every symbol in a
  configuration is preceded by a bit string, which represents a {\em
    counter}.  By (\ref{U0-2}) every counter is non-empty, the first
  symbol in a configuration is preceded by a counter from $0^+$, the
  last symbol is preceded by a counter from $1^+$. Moreover, by
  (\ref{U0-3}), the first configuration is an initial configuration,
  whereas by (\ref{U0-4}), the last configuration is accepting
  (i.e. the current state is $q_f$).

  For the further considerations, let us fix some $x \in U_0$. Hence,
  we can factorize $x$ as $x = x_1 \# x_2 \# \cdots x_n\#$ such that:
  \begin{itemize}
  \item For every $1 \leq i \leq n$, there exist $m_i \geq 1$, 
  $a_{i,0}, \ldots, a_{i,m_i} \in \Omega$ and counters
  $u_{i,0}, \ldots, u_{i,m_i} \in \{0,1\}^+$ such that
  $x_i = \prod_{j=0}^{m_i} u_{i,j} a_{i,j}$.
  \item For every $1 \leq i \leq n$,  
  $u_{i,0} \in 0^+$, $u_{i,m_i} \in 1^+$, and
  $\pi(x_i) \in \Gamma^* Q \Gamma^+$.
  \item $\pi(x_1) \in q_0 \Sigma^* \Box^*$ and 
  $\pi(x_n) \in \Gamma^* q_f \Gamma^+$
  \end{itemize}
  We next want to construct, from $m\in\dN$, a small formula
  expressing that $x$ is an accepting $2^m$-computation. To achieve
  this, we add some structure around strings
  from~$U_0$. Then the formula we are seeking has to ensure two facts:
  \begin{enumerate}[(a)]
  \item The counters behave correctly, i.e. for all $1 \leq i \leq n$
    and $0 \leq j \leq m_i$, we have $|u_{i,j} | = m$ and if $j < m_i$,
    then $\val(u_{i,j+1}) = \val(u_{i,j})+1$.  Note that this enforces
    $m_i = 2^m-1$ for all $1 \leq i \leq n$.
  \item For two successive configurations, the second one is the
    successor configuration of the first one with respect to the machine
    $M$, i.e., $\pi(x_i) \vdash_M \pi(x_{i+1})$ for all $1 \leq i < n$.
  \end{enumerate}
  In order to achieve (a), we introduce the following three binary
  relations; it is straightforward to exhibit 2-dimensional automata for these
  relations:
  \begin{align*}
  \delta \ =\ & \{ (w,\, w \otimes w) \mid w \in U_0 \} \\
  \sigma_0 \ = \ & \{  \bigl( (0 v_1 \# 0 v_2 \# \cdots 0 v_n\#) \otimes w, \, 
  (v_1 0 \# v_2 0 \# \cdots v_n 0\#) \otimes w \bigr) \mid \\
  & \qquad\qquad\qquad\qquad\qquad\qquad\qquad\qquad w \in U_0, v_1,\ldots,v_n \in
  (\Delta\setminus\{\#\})^* \}  \\
  \sigma_\Omega \ = \ & \{  \bigl( (a_1 v_1 \# a_2 v_2 \# \cdots a_n v_n\#) \otimes w, \, 
  (v_1 a_1 \# v_2 a_2 \# \cdots v_n a_n\#) \otimes w \bigr) \mid \\
  & \qquad\qquad\qquad\qquad\qquad\qquad\qquad\qquad w \in U_0, 
  a_1,\ldots,a_n \in \Omega, v_1,\ldots,v_n \in (\Delta\setminus\{\#\})^* \} 
  \end{align*}
  Hence, $\delta$ just duplicates a string from $U_0$ and $\sigma_0$
  cyclically rotates every configuration to the left for one symbol, provided the
  first symbol is $0$, whereas $\sigma_\Omega$ rotates symbols from
  $\Omega$.  Moreover, let $U_1$ be the following regular language
  over $\Delta^* \otimes \Delta^*$:
  $$
  U_1 \ = \ \biggr(\bigl(\{ u \otimes v \mid u, v \in \{0,1\}^+, |u|=|v|, \val(u) = \val(v)+1
  \ \modulo \ 2^{|u|} \} (\Omega
  \times \Omega) \bigr)^+ (\#,\#) \biggl)^+
  $$
  Clearly, $U_1$ is a regular language.  The crucial fact is the
  following:

  \medskip

  \noindent {\bf Fact 1.} For every $m \in \dN$, the following two
  properties are equivalent (recall that $x \in U_0$):
  \begin{itemize}
  \item There exist
  $y_1, y_2, y_3 \in \Delta^* \otimes \Delta^*$ such that
  $\delta(x,y_1)$, $\sigma_0^{m}(y_1,y_2)$, $\sigma_\Omega(y_2,y_3)$, $y_3 \in U_1$.
  \item For all
  $1 \leq i \leq n$ and $0 \leq j \leq m_i$, we have 
  $|u_{i,j}| = m$ and if $j < m_i$, then 
  $\val(u_{i,j+1}) = \val(u_{i,j})+1$. 
  \end{itemize}
  Assume now that $x \in U_0$ satisfies one (and hence both) of the
  two properties from Fact 1 for some $m$. It follows that $m_i = 2^m-1$
  for all $1 \leq i \leq n$ and
  \begin{equation}\label{string x}
  x= x_1 \# x_2 \# \cdots x_n \#, \text{ where } x_i = \prod_{j=0}^{2^m-1} \bin_m(j)
  a_{i,j} \text{ for every } 1 \leq i \leq n\ .
  \end{equation}
  In order to establish (b) we need additional structure. The idea is,
  for every counter value~$0\le j<2^m$, to have a word $y_j$ that
  coincides with $x$, but has all the occurrences of $\bin_m(j)$
  marked. Then an automaton can check that successive occurrences of
  the counter $\bin_m(j)$ obey the transition condition of the Turing
  machine. There are two problems with this approach: first, in order
  to relate $x$ and $y_j$, we would need a binary relation of
  degree~$2^m$ (for arbitrary $m$) and, secondly, an automaton cannot
  mark all the occurrences of $\bin_m(j)$ at once (for some $j$). 
  In order to solve these problems, we introduce a
  binary relation $\mu$, which for every $x \in U_0$ as in
  (\ref{string x}) generates a binary tree of depth~$m$ with root $x$;
  this will be the only relation in our string automatic structure that
  causes exponential growth. This relation will mark in $x$ every
  occurrence of an arbitrary counter. For this, we need two copies
  $\ol 0$ and $\ul 0$ of $0$ as well as two copies $\ol 1$ and $\ul 1$
  of $1$. For $b\in\{0,1\}$, define the mapping
  \[
    f_b : \{ \ul{0}, \ol{0}, \ul{1}, \ol{1} \}^* \{0,1\}^+ \to  
          \{ \ul{0}, \ol{0}, \ul{1}, \ol{1} \}^+ \{0,1\}^*
  \]
  as follows (where $u \in \{ \ul{0}, \ol{0}, \ul{1}, \ol{1} \}^*$, $c
  \in \{0,1\}$, and $v \in \{0,1\}^*$):
  \[
    f_b(u c v) = 
     \begin{cases}
       u \ul{c} v & \text{if $b \neq c$} \\
       u \ol{c}  v & \text{if $b = c$}
     \end{cases}
  \]
  We extend $f_b$ to a function on $((\{\ul{0}, \ol{0}, \ul{1}, \ol{1}
  \}^*\{0,1\}^+\Omega)^+ \#)^*$ as follows: Let $w = w_1 a_1 \cdots
  w_\ell a_\ell$ with $w_i \in \{\ul{0}, \ol{0}, \ul{1}, \ol{1}
  \}^*\{0,1\}^+$ and $a_i \in \Omega \cup \Omega\#$.  Then $f_b(w) =
  f_b(w_1) a_1 \cdots f_b(w_\ell) a_\ell$; this mapping can be computed with
  a synchronized transducer. Hence, the relation
  \[
    \mu = f_0\cup f_1
        = \{(u,f_b(u)) \mid 
               u\in((\{\ul 0,\ol 0,\ul 1,\ol 1\}^*\{0,1\}^+\Omega)^+\#)^*, 
               b\in\{0,1\}\}
  \] 
  can be recognized by a 2-dimensional automaton.

  Let $x\in U_0$ as in (\ref{string x}), let the word $y$ be obtained
  from $x$ by overlining or underlining each bit in $x$, and let
  $u\in\{0,1\}^m$ be some counter. We say \emph{the counter $u$ is
    marked in $y$} if every occurrence of the counter $u$ is marked by
  overlining each bit, whereas all other counters contain at least one
  underlined bit.

  \medskip

  \noindent {\bf Fact 2.} Let $x \in U_0$ be as in (\ref{string x}).
  \begin{itemize}
  \item For all counters $u\in\{0,1\}^m$, there exists a unique
    word~$y$ with $(x,y)\in\mu^m$ such that the counter $u$ is marked
    in~$y$.
  \item If $(x,y)\in\mu^m$, then there exists a unique counter $u \in
    \{0,1\}^m$ such that $u$ is marked in $y$.
  \end{itemize}
  Now, we can achieve our final goal, namely checking whether two
  successive configurations in $x \in U_0$ represent a transition of
  the machine $M$. Let the counter $u \in \{0,1\}^m$ be marked in~$y$.
  We describe a finite automaton $A_2$ that checks on the string~$y$,
  whether at position $\val(u)$ successive configurations in~$x$ are
  ``locally consistent''.  The automaton $A_2$ searches for the first
  marked counter in~$y$. Then it stores the next three symbols $a_1,
  a_2, a_3$ from $\Omega$ (if the separator~$\#$ is seen before, then
  only one or two symbols may be stored), walks right until it finds
  the next marked counter, reads the next three symbols~$b_1, b_2,
  b_3$ from $\Omega$, and checks whether $(a_1a_2a_3,
  b_1b_2b_3) \in \alpha_M$, where $\alpha_M$ is from (\ref{equiv alpha}).  
  If this is not the case, the automaton
  will reject, otherwise it will store $b_1b_2b_3$ and repeat the
  procedure described above.  Let $U_2 = L(A_2)$. Together with Fact~1
  and 2, the behavior of $A_2$ implies that for all $x \in U_0$ and
  all $m \in \dN$, $x$ represents an accepting $2^m$-computation of
  $M$ if and only if
  $$
   \exists y_1, y_2, y_3 \ \biggl( \delta(x,y_1) \ \wedge \ \sigma_0^{m}(y_1,y_2)  \ \wedge \ \sigma_\Omega(y_2,y_3) 
   \ \wedge \ y_3 \in U_1 \biggr) \ \wedge \
   \forall y \ \biggl( \mu^m(x,y) \ \to \ y \in U_2  \biggr)\ .
   $$
  Let us now fix some input $w = a_1 a_2 \cdots a_n \in \Sigma^*$ with
  $|w|=n$, and let $a_{n+1} = \Box$ and $m = 2^n$.  Thus, $w$ is
  accepted by $M$ if and only if there exists an accepting
  $2^m$-computation~$x$ such that in the first configuration of $x$,
  the tape content is of the form $w \Box^+$.  It remains to add some
  structure that allows us to express the latter by a
  formula. But this is straightforward: 
  Let $\triangleright$ be a new symbol and let 
  $\Pi = \Delta  \cup \{\ul{0}, \ol{0}, \ul{1}, \ol{1}, \triangleright\}$;
  this is our final alphabet.  Define the binary relations
  $\iota_{0,1}$ and $\iota_a$ ($a \in \Omega$) as follows:
  \begin{align*}
  \iota_{0,1} \ =\ & \{ (u \triangleright a v, u a \triangleright v) \mid 
  a \in \{0,1\}, u,v \in \Delta^*, 
  u a v \in U_0 \} \ \cup \ \{ (0 v, 0 \triangleright v) \mid  v \in
  \Delta^*, 0v \in U_0 \} \\
   \iota_a \ =\ & \{ (u \triangleright a v, u a \triangleright v) \mid u,v \in \Delta^*, u a v \in U_0 \} \ . 
  \end{align*}
  Then,
  $\mcA = (\Pi^* \cup (\Pi^* \otimes \Pi^*), \delta, \sigma_0, \sigma_\Omega,
  \mu, \iota_{0,1}, (\iota_a)_{a \in \Omega}, U_0, U_1, U_2)$
  is a string automatic structure of bounded degree
  such that $w$ is accepted by $M$ if and
  only if the following formula is true in $\mcA$:
  \[
  \exists x \in U_0 \left\{ 
  \begin{array}{l}
   \exists y_1, y_2, y_3 \ \biggl( \delta(x,y_1) \ \wedge \ \sigma_0^{m}(y_1,y_2)  \ \wedge \ \sigma_\Omega(y_2,y_3) 
   \ \wedge \ y_3 \in U_1 \biggr) \ \wedge \\
   \forall y \ \biggl( \mu^m(x,y) \ \to \  y \in U_2   \biggr) \ \wedge
   \\ 
   \displaystyle
   \exists y_0,z_0, \ldots, y_{n+1},z_{n+1} \biggl(  \iota_{0,1}^m(x,y_0) \
   \wedge \ \iota_{q_0}(y_0,z_0) \ \wedge \ \bigwedge_{i=1}^{n+1}
   \iota_{0,1}^m(z_{i-1},y_i) \, \wedge \, 
   \iota_{a_i}(y_i,z_i) \biggr)
  \end{array}\right\}
  \]
  By Lemma~\ref{L long paths} we can compute in time $O(\log(m)) =
  O(n)$ an equivalent formula over the signature of $\mcA$.  This
  concludes the proof.  \qed
\end{proof}

The following theorem, which proves an analogous result for
tree automatic structures, uses alternating Turing machines, see
\cite{ChaKS81,Pap94} for more details. Roughly speaking, an
\emph{alternating Turing machine} is a nondeterministic Turing
machine, where the set of states is partitioned into accepting,
existential, and universal states. A configuration is accepting, if
either (i) the current state is accepting, or (ii) the current state
is existential and at least one successor configuration is accepting,
or (iii) the current state is universal and every successor
configuration is accepting.  By \cite{ChaKS81}, $k${\sf EXPTIME} is the set
of all problems that can be accepted in space $\exp(k-1,n^{O(1)})$ on an
alternating Turing machine (for all $k \geq 1$).

\begin{theorem} \label{T 3EXPTIME} There exists a fixed tree automatic
  structure $\mcA$ of bounded degree such that $\FOTh(\mcA)$ is
  {\sf 3EXPTIME}-hard.
\end{theorem}

\begin{proof}
Let $M$ be a fixed {\em alternating} Turing machine with a 
space bound of $\exp(2,n)$ such that $M$ accepts a {\sf 3EXPTIME}-complete language.
W.l.o.g. every configuration, where the current state is either existential
or universal has exactly two successor configurations. 
Let $\Sigma$, $\Gamma$, $Q$, and $\Omega$ have the same meaning as in the
previous proof. Moreover, let $\Delta = \Omega \cup \{ 0,1, \#_\exists, \#_\forall \}$.

The idea is that a binary tree $x$ over the alphabet $\Delta$ can 
encode a computation tree for some input. Configurations can be encoded
by linear chains over the alphabet $\Omega \cup \{0,1\}$ as in the previous proof.
The separator symbol 
$\#_\exists$ is used to separate an existential configuration from 
a successor configuration, whereas the
separator symbol $\#_\forall$ is used to separate a universal configuration from 
its two successor configurations. Hence, a $\#_\exists$-labeled node has 
exactly one child, whereas a $\#_\forall$-labeled node has 
exactly two children.
Checking whether the counters 
behave correctly can be done similarly to the previous proof by introducing
binary relations $\sigma_0$ and $\sigma_\Omega$, which rotate symbols within
configurations. Remember that in our tree encoding, configurations are just
long chains. Also the marking of some specific counter
can be done in the same way as before. Finally, having marked
some specific counter allows to check with a top-down tree automaton, 
whether the tree $x$ represents
indeed  a valid computation tree. Of course, the tree automaton 
has to check whether the current configuration is existential or universal.
In case of a universal configuration, the automaton branches at the next
separator symbol $\#_\forall$. If e.g. the current configuration is universal but 
the next separator symbol is $\#_\exists$, then the automaton rejects the tree. 
\qed
\end{proof}

The proof of the next result is in fact a simplification of 
the proof of Theorem~\ref{T 2EXPSPACE}, since we do not need
counters.

\begin{theorem} \label{T EXPSPACE}
There exists a fixed string automatic structure $\mcA$ of bounded degree
and polynomial growth (in fact linear growth)
such that $\FOTh(\mcA)$ is {\sf EXPSPACE}-hard.
\end{theorem}

\begin{proof}
Let $M$ be a fixed Turing machine with a
space bound of $2^n$ such that $M$ accepts an {\sf EXPSPACE}-complete language.
Let $\Sigma$, $\Gamma$, $Q$, $q_0$, $q_f$, $\Box$, and $\Omega$ have
the usual meaning.
Let $\Delta = \{\#\} \cup \Omega$.
This time, for $m \in \dN$, an {\em accepting $m$-computation} is
a string $x_1 \# x_2 \# \cdots x_n \#$, where 
$x_1, \ldots, x_n \in \Gamma^* Q \Gamma^+$
are configurations with $|x_i| = m$ ($1 \leq i \leq n$),
$x_i \vdash_M x_{i+1}$ ($1 \leq i < n$), $x_1 \in q_0 \Sigma^* \Box^*$, and
$x_n \in \Gamma^* q_f \Gamma^+$.
Let $U_0$ be the fixed regular language 
$$
U_0 \ = \  (\Gamma^* Q \Gamma^+\#)^+ \ \cap \
 q_0 \Sigma^* \Box^* \# \Delta^* \ \cap \ \Delta^* q_f (\Delta \setminus
 \{\#\})^* \#\ .
$$
The following binary relations $\delta$ and $\sigma_\Omega$
can be easily recognized by 2-dimensional automata:
\begin{align*}
\delta \ =\ & \{ (w,\, w \otimes w) \mid w \in U_0 \} \\
\sigma_\Omega \ = \ & \{ (av \otimes w, \, 
va \otimes w) \mid w \in U_0, a \in \Omega, v \in \Delta^* \}  
\end{align*}
Moreover, let $U_1$ be the following regular language over $\Delta^* \otimes
\Delta^*$:
$$
U_1 \ = \ \{ \# u \otimes v\# \mid u, v \in \Omega^+, |u|=|v|, v \vdash_M
u\}^+ \{ \# u \otimes v\# \mid u, v \in \Omega^+, |u|=|v|\} \ .
$$
Then, for every $x \in U_0$ and $m \in \dN$ we have: $x$ is an
accepting $m$-computation if and only if there exist $y_1, y_2
\in \Delta^* \otimes \Delta^*$ such that $\delta(x,y_1)$,
$\sigma_\Omega^{m}(y_1,y_2)$, and $y_2 \in U_1$.

Let us now fix some input $w = a_1 \cdots a_n \in \Sigma^*$ with $|w|=n$,
let $a_{n+1} = \Box$, and let $m = 2^n$.
Thus, $w$ is accepted by $M$ if and only if there exists 
an accepting $m$-computation $x$ such that in the first configuration 
of $x$, the tape content is of the form $w \Box^+$.
It remains to add some structure that allows us 
to express the latter by a formula.
This can be done similarly to the proof of Theorem~\ref{T 2EXPSPACE}: 
Let  $\Pi = \Delta \cup \{\triangleright\}$, where $\triangleright$ is a new symbol and
define the binary relations $\iota_a$ 
($a \in \Sigma \cup \{\Box\}$) as follows: 
$$
\iota_a \ =\ 
\{ (q_0 a v, q_0 a \triangleright v) \mid  v \in
\Delta^*, q_0 a v \in U_0 \} \ \cup \
 \{ (u \triangleright a v, u a \triangleright v) \mid u,v \in
\Delta^*, u a v \in U_0 \}
$$
Then, 
$\mcA = ( \Pi^* \cup (\Delta^* \otimes \Delta^*), \delta,
\sigma_\Omega, (\iota_a)_{a \in \Sigma \cup \{\Box\}}, U_0, U_1)$
is a fixed string automatic structure of bounded degree and linear growth.
For the latter note that the Gaifman graph of $\mcA$ is just a
disjoint union of cycles and finite paths (in fact, every node has degree
at most 2). 
Moreover, $w$ is accepted by $M$ if
and only if the following statement is true in $\mcA$:
\begin{equation} \label{reduction formula expspace}
\exists x \in U_0 \left\{ 
\begin{array}{l}
 \exists y_1, y_2 \ \biggl( \delta(x,y_1) \ \wedge \ \sigma_\Omega^{m}(y_1,y_2)  
 \ \wedge \ y_2 \in U_1 \biggr) \ \wedge \\ \displaystyle
 \exists y_0, \ldots, y_n \biggl( \iota_{a_1}(x,y_0)  \ \wedge \
  \bigwedge_{i=1}^n
 \iota_{a_i}(y_{i-1},y_i)  \biggr)
\end{array}\right\} \ .
\end{equation}
By Lemma~\ref{L long paths} this concludes the proof.
\qed
\end{proof}

The next result can be easily shown by combining the techniques from the proof
of Theorem~\ref{T 3EXPTIME} and \ref{T EXPSPACE}. We leave the details for the reader.

\begin{theorem}
There exists a fixed tree automatic structure $\mcA$ of bounded degree
and polynomial growth (in fact linear growth)
such that $\FOTh(\mcA)$ is {\sf 2EXPTIME}-hard.
\end{theorem}

\section{Bounded quantifier alternation depth}

In this section we prove some facts about first-order fragments of
fixed quantifier alternation depth. These results will follow easily
from the constructions in the preceding section.

For $n \geq 0$, $\Sigma_n$-formulas and $\Pi_n$-formulas are
inductively defined as follows:
\begin{itemize}
\item A quantifier-free first-order formula is a $\Sigma_0$-formula
as well as a $\Pi_0$-formula.
\item If $\varphi(x_1,\ldots,x_n,y_1,\ldots,y_m)$ 
is a $\Sigma_n$-formula, then
$\forall x_1 \cdots \forall x_n : \varphi(x_1,\ldots,x_n,y_1,\ldots,y_m)$
is a $\Pi_{n+1}$-formula.
\item If $\varphi(x_1,\ldots,x_n,y_1,\ldots,y_m)$ 
is a $\Pi_n$-formula, then
$\exists x_1 \cdots \exists x_n : \varphi(x_1,\ldots,x_n,y_1,\ldots,y_m)$
is a $\Sigma_{n+1}$-formula.
\end{itemize}
The $\Sigma_n$-theory $\Sigma_n$-$\FOTh(\mcA)$ of a structure $\mcA$ is
the set of all $\Sigma_n$-formulas in $\FOTh(\mcA)$; the
$\Pi_n$-theory is defined analogously. For a class $\dC$ of tree
automatic presentations, the \emph{$\Sigma_n$-model checking problem
  $\Sigma_n$-$\Decision(\dC)$ of $\dC$} denotes the set of all pairs
$(P,\varphi)$ where $P \in \dC$, and $\varphi \in \Sigma_n$-$\FOTh(\mcA(P))$.

The following result can be found in \cite{BluG00}:

\begin{theorem}[cf.\ \cite{BluG00}]\label{T BluG00} 
  The $\Sigma_1$-model checking problem
  $\Sigma_1$-$\Decision(\mathsf{SA})$ for all string automatic
  presentations is in {\sf PSPACE}.
  Moreover, there is a fixed string automatic structure
  with a {\sf PSPACE}-complete $\Sigma_1$-theory.
\end{theorem}

{}From our construction in the proof of Theorem~\ref{T EXPSPACE},
we can slightly sharpen the lower bound in this theorem:

\begin{theorem}
  There exists a fixed string automatic structure of bounded degree
  and linear growth with a {\sf PSPACE}-complete $\Sigma_1$-theory.
\end{theorem}

\begin{proof}
Take the structure $\mcA$ from the proof of Theorem~\ref{T EXPSPACE} and let 
$M$ be a fixed linear bounded automaton with a {\sf PSPACE}-complete
acceptance problem. If we replace the number $m$ in the formula
(\ref{reduction formula expspace}) by the input length 
$n$, then (\ref{reduction formula expspace}) is equivalent
to the following formula, which is equivalent to a $\Sigma_1$-formula:
$$
\exists x \in U_0 \left\{ 
\begin{array}{l} \displaystyle
 \exists y_0, \ldots, y_{n+1} \ \biggl( \delta(x,y_0) \ \wedge \
 \bigwedge_{i=0}^n \sigma_\Omega(y_i,y_{i+1})  
 \ \wedge \ y_{n+1} \in U_1 \biggr) \ \wedge \\[5mm] \displaystyle
 \exists y_1, \ldots, y_n \biggl( \iota_{a_1}(x,y_1)  \ \wedge \
  \bigwedge_{i=2}^n
 \iota_{a_i}(y_{i-1},y_i)  \biggr)
\end{array}\right\} \ .
$$
This formula is true in $\mcA$ if and only if the linear
bounded automaton accepts the input $w = a_1 \cdots a_n$.
\qed
\end{proof}

Let us now move on to $\Sigma_2$-formulas and structures of arbitrary growth:

\begin{theorem}\label{T Sigma2} 
  The $\Sigma_2$-model checking problem
  $\Sigma_2$-$\Decision(\mathsf{SA})$ for all string automatic
  presentations is in {\sf EXPSPACE}.
  Moreover, there is a string automatic structure of bounded
  degree with an {\sf EXPSPACE}-complete $\Sigma_2$-theory.
\end{theorem}

\begin{proof}
  For the upper bound, let $P$ be a string automatic presentations of
  the automatic structure $\mcA(P)=\mcA$ and let
  $$
  \psi = \exists x_1 \cdots \exists x_n \forall y_1 \cdots \forall y_m :
  \varphi
  $$ 
  be a $\Sigma_2$-sentence. The sentence $\psi$ is equivalent to
  $$
  \exists x_1 \cdots \exists x_n \neg \exists y_1 \cdots \exists y_m :
  \neg \varphi\ .
  $$
  Negations in $\neg \varphi$ can be moved down to the level of atomic
  predicates.  Then, we can built an $(n+m)$-dimensional automaton for
  $\neg \varphi$ with $\exp(1,|\psi|^{O(1)})$ many states. Projection
  onto the tracks corresponding to the variables $x_1,\ldots,x_n$
  results again into an automaton with $\exp(1,|\psi|^{O(1)})$ many
  states.  Hence, for $\neg \exists y_1 \cdots \exists y_m : \neg
  \varphi$ there exists an $n$-dimensional automaton with
  $\exp(2,|\psi|^{O(1)})$ many states. But, we do not need to
  construct this automaton explicitly but only have to check emptiness
  of its language, which can be done on the fly in exponential space.

  For the lower bound, we reuse our construction from the proof of
  Theorem~\ref{T 2EXPSPACE}. We start with an
  $\exp(1,n)$-space-bounded machine $M$ that accepts an {\sf
    EXPSPACE}-complete language. We carry out the same construction as
  in the proof of Theorem~\ref{T 2EXPSPACE}, but replace $2^m$
  (resp. $m$) everywhere by $m$ (resp. the input length $n$). In
  addition, we need the following (trivial) analogue of Lemma~\ref{L
    long paths}: Let the signature $\mcS$ contain a binary symbol $r$.
  {}From a given number $n$ (encoded unary), we can construct in
  linear time a $\Sigma_1$-formula $\varphi^{(n)}(x,y)$ such that for
  every $\mcS$-structure $\mcA$ and all elements $a,b \in \mcA$ we
  have: $(a,b) \in r^{n}$ if and only if $\mcA \models
  \varphi^{(n)}(a,b)$.

  Then, the final formula from the proof of Theorem~\ref{T 2EXPSPACE}
  can be written as
  \[
  \exists x \in U_0 \left\{ 
  \begin{array}{l}
   \exists y_1, y_2, y_3 \ \biggl( \delta(x,y_1) \ \wedge \ \sigma_0^{(n)}(y_1,y_2)  \ \wedge \ \sigma_\Omega(y_2,y_3) 
   \ \wedge \ y_3 \in U_1 \biggr) \ \wedge \\
   \forall y \ \biggl( \neg\mu^{(n)}(x,y) \ \lor \  y \in U_2   \biggr) \ \wedge
   \\ 
   \displaystyle
   \exists y_0,z_0, \ldots, y_{n+1},z_{n+1} \biggl(  \iota_{0,1}^{(n)}(x,y_0) \
   \wedge \ \iota_{q_0}(y_0,z_0) \ \wedge \ \bigwedge_{i=1}^{n+1}
   \iota_{0,1}^{(n)}(z_{i-1},y_i) \, \wedge \, 
   \iota_{a_i}(y_i,z_i) \biggr)
  \end{array}\right\}.
  \]
  This formula is equivalent to a $\Sigma_2$-formula. Moreover, this 
  formula is true in the string automatic structure $\mcA$ (of bounded degree)
  from the proof of Theorem~\ref{T 2EXPSPACE}, if and only if the input $w = a_1
  a_2 \cdots a_n$ is accepted by the machine $M$.
  \qed
\end{proof}

As before, Theorems~\ref{T BluG00}--\ref{T Sigma2} can be extended to
tree automatic structures as follows:

\begin{theorem}
The following holds:
  \begin{enumerate}
  \item The $\Sigma_1$-model checking problem
    $\Sigma_1$-$\Decision(\mathsf{TA})$ for all tree automatic
    presentations is in {\sf EXPTIME}.
  \item There exists a fixed tree automatic structure of bounded
    degree and linear growth with an {\sf EXPTIME}-complete
    $\Sigma_1$-theory.
  \item The $\Sigma_2$-model checking problem
    $\Sigma_2$-$\Decision(\mathsf{TA})$ for all tree automatic
    presentations is in {\sf 2EXPTIME}.
  \item There exists a tree automatic structure of bounded
    degree with a {\sf 2EXPTIME}-complete $\Sigma_2$-theory.
  \end{enumerate}
\end{theorem}

\section{Open problems}

The most obvious open question regards the uniform first-order theory for
(injective) tree automatic structures: we do not know whether it is
{\sf 4EXPTIME}-hard. Moreover, we don't know
an upper bound for the uniform first-order theory
for arbitrary tree automatic structures. The reason is that we do not
know the complexity of transforming such a presentation into an
equivalent injective one (which is possible by~\cite{ColL07}).

In \cite{BluG00,KhoRS04}, it is shown that not only the
first-order theory of every string automatic structure is (uniformly)
decidable, but even its extension by the quantifiers ``there are
infinitely many $x$ with \dots'' and ``the number of $x$ satisfying
\dots is divisible by $p$''. In \cite{KusL08}, we proved that this
extended theory can be decided in triply exponential time for
($\omega$)-automatic structures of bounded degree. It is not clear
whether the doubly-exponential upper bound proved in this paper
extends to this more expressive theory.

Recall that there are tree automatic structures which are not string
automatic. Provided {\sf 2EXPSPACE} $\neq$ {\sf 3EXPTIME}, our results on the
non-uniform first-order theories imply the existence of such a
structure of bounded degree (namely the tree automatic structure
constructed in the proof of Theorem~\ref{T 3EXPTIME}). But no example
is known that does not rest on the complexity theoretic assumption
{\sf 2EXPSPACE} $\neq$ {\sf 3EXPTIME}.

For $n \geq 3$, the precise complexity of the $\Sigma_n$-theory of a
string/tree automatic structure of bounded degree remains open. We
know that these theories belong to {\sf 2EXPSPACE} for string
automatic structures and to {\sf 3EXPTIME} for tree automatic
structures. Moreover, from our results for the $\Sigma_2$-fragment we
obtain lower bounds of {\sf EXPSPACE} and {\sf 2EXPTIME},
respectively.
\begin{conjecture}
  For every $n\geq3$, the problems 
  $\Sigma_n$-$\Decision(\mathsf{SAb})$ and
  $\Sigma_n$-$\Decision(\mathsf{TAb})$ 
   belong to {\sf EXPSPACE} and {\sf 2EXPTIME}, respectively.
\end{conjecture}
A possible attack to this conjecture would follow the line of argument
in the proof of Theorem~\ref{T-upper} and would therefore be based on
Gaifman's theorem. To make this work, the exponential bound in
Gaifman's theorem would have to be reduced which leads to the
following conjecture.
\begin{conjecture}
  Let $\mcA$ be a structure, $(a_1,\ldots,a_k), (b_1,\ldots,b_k) \in
  \mcA^k$, $d \geq 0$, and $D_1, \ldots, D_k \geq d\cdot 2^n$ such that
  \begin{equation*}
    (\mcA \rest (\bigcup_{i=1}^k S(D_i,a_i) ), \ a_1, \ldots, a_k) \simeq
    (\mcA \rest (\bigcup_{i=1}^k S(D_i,b_i) ), \ b_1, \ldots, b_k)\ .
  \end{equation*}
  Then, for every $\Sigma_n$-formula $\varphi(x_1,\ldots,x_k)$ of
  quantifier depth at most $d$, we
  have: $$\mcA\models\varphi(a_1,\ldots,a_k) \ \Longleftrightarrow \
  \mcA\models\varphi(b_1,\ldots,b_k) \ .$$

\end{conjecture}

\bibliographystyle{abbrv}

\end{document}